\pdfoutput=1
\documentclass[fleqn,usenatbib,usedcolumn]{mnras}

\usepackage{graphicx,amssymb,hyperref}
\usepackage{multirow}
\usepackage{color}

\usepackage[fleqn]{amsmath}
\usepackage{abbrev}
\usepackage{siunitx}
\usepackage{comment} 
\usepackage{gensymb}
\usepackage{booktabs,tabularx}
\usepackage{flushend}
\usepackage{xspace}
\usepackage{physics}
\usepackage{xcolor}

\usepackage{nicefrac}
\usepackage{units}
\usepackage{verbatim}
\usepackage[shortlabels]{enumitem}

\setlist[itemize]{leftmargin=5pt}

\def\mnras{MNRAS}
\def\apj{ApJ}

\def\simlt{\lower.5ex\hbox{$\; \buildrel < \over \sim \;$}}
\def\simgt{\lower.5ex\hbox{$\; \buildrel > \over \sim \;$}}
\def\etal{{\it et al.}}

\def\D{\mathrm{d}}

\def\kpc{\mathrm{\, kpc}}

\def\mpc{\mathrm{\, Mpc}}
\def\gev{\mathrm{\, GeV}}
\def\kev{\mathrm{\, keV}}
\def\msun{\mathrm{\, M_\odot}}
\def\kms{\mathrm{\, km \, s^{-1}}}
\def\cmsg{\, \mathrm{cm^2 \, g^{-1}}}

\def\effein{\theta_\mathrm{E,eff}}

\newcommand{\eagle}{\textsc{eagle}\xspace}
\newcommand{\ceagle}{\textsc{c-eagle}\xspace}

\newcommand{\bahamas}{\textsc{bahamas}\xspace}

\def\gs{\mathrel{\raise1.16pt\hbox{$>$}\kern-7.0pt \lower3.06pt\hbox{{$\scriptstyle \sim$}}}}         
\def\ls{\mathrel{\raise1.16pt\hbox{$<$}\kern-7.0pt \lower3.06pt\hbox{{$\scriptstyle \sim$}}}}   

\newcommand{\vect}[1]{\boldsymbol{#1}}
\newcommand{\code}[1]{\texttt{#1}}

\newcommand{\be}{\begin{equation}}
\newcommand{\ee}{\end{equation}}
\newcommand{\ba}{\begin{eqnarray}}
\newcommand{\ea}{\end{eqnarray}}

\newcommand{\orcid}[1]{\href{https://orcid.org/#1}{\includegraphics[scale=0.08]{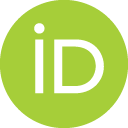}}}

\DeclareGraphicsExtensions{.pdf,.png}

\title[BAHAMAS-SIDM]{Observable tests of self-interacting dark matter in galaxy clusters: cosmological simulations with SIDM and baryons}

\author[A.\ Robertson \etal]
{\parbox{\textwidth}{Andrew Robertson$^1$\thanks{e-mail: {\tt andrew.robertson@durham.ac.uk}}\orcid{0000-0002-0086-0524},
David Harvey$^2$\orcid{0000-0002-6066-6707}, Richard Massey$^{1,3}$\orcid{0000-0002-6085-3780}, Vincent Eke$^1$\orcid{0000-0001-5416-8675}, Ian G. McCarthy$^4$\orcid{0000-0002-1286-483X}, Mathilde Jauzac$^{1,3,5}$\orcid{0000-0003-1974-8732}, Baojiu Li$^1$\orcid{0000-0002-1098-9188} and Joop Schaye$^6$\orcid{0000-0002-0668-5560}}
\vspace{0.3cm}
\\$^1$Institute for Computational Cosmology, Durham University, South Road, Durham DH1 3LE, UK
\\$^2$Lorentz Institute, Leiden University, Niels Bohrweg 2, Leiden, NL-2333 CA, the Netherlands
\\$^3$Centre for Extragalactic Astronomy, Department of Physics, Durham University, Durham DH1 3LE, UK
\\$^4$Astrophysics Research Institute, Liverpool John Moores University, 146 Brownlow Hill, Liverpool L3 5RF, UK
\\$^5$Astrophysics and Cosmology Research Unit, School of Mathematical Sciences, University of KwaZulu-Natal, Durban 4041, South Africa
\\$^6$Leiden Observatory, Leiden University, PO Box 9513, NL-2300 RA Leiden, the Netherlands
}

\begin{document}

\maketitle

\label{firstpage}

\begin{abstract}

We present BAHAMAS-SIDM, the first large-volume, $(400 \, h^{-1} \mpc)^{3}$, cosmological simulations including both self-interacting dark matter (SIDM) and baryonic physics.
These simulations are important for two primary reasons: 1) they include the effects of baryons on the dark matter distribution 2) the baryon particles can be used to make mock observables that can be compared directly with observations. As is well known, SIDM haloes are systematically less dense in their centres, and rounder, than CDM haloes. Here we find that that these changes are not reflected in the distribution of gas or stars within galaxy clusters, or in their X-ray luminosities. However, gravitational lensing observables can discriminate between DM models, and we present a menu of tests that future surveys could use to measure the SIDM interaction strength. We ray-trace our simulated galaxy clusters to produce strong lensing maps. Including baryons boosts the lensing strength of clusters that produce no critical curves in SIDM-only simulations. Comparing the Einstein radii of our simulated clusters with those observed in the  CLASH survey, we find that at velocities around $1000 \kms$ an SIDM cross-section of $\sigma/m \gtrsim 1 \cmsg$ is likely incompatible with observed cluster lensing.

\end{abstract}


\section{Introduction}

Self-interacting dark matter (SIDM) has become an attractive alternative to collisionless cold dark matter (CDM) because it can alleviate tensions between the results of DM-only simulations, and observations of dwarf and low-mass galaxies \citep{2000PhRvL..84.3760S, 2012MNRAS.423.3740V, 2013MNRAS.430...81R, 2013MNRAS.431L..20Z, 2015MNRAS.453...29E, 2016PhRvL.116d1302K, 2017PhRvL.119k1102K, 2017MNRAS.468.2283C}. These tensions arise from the low inferred DM densities at the centre of some observed galaxies, which are at odds with the CDM-only prediction of steeply-rising central density profiles \citep[for a review see][]{2015PNAS..11212249W}. The extent to which these tensions are indications of new physics and not simply the result of neglecting (or improperly treating) baryonic physics when making the theoretical predictions is hotly debated. For example, DM haloes can be kinematically heated by rapid fluctuations to the gravitational potential, which could be produced by feedback from stars driving gas out of galaxies \citep[see][for a review]{2014Natur.506..171P}. This heating lowers the central density of DM haloes in an analogous manner to the heating of DM particles through self-interactions, though recent evidence that dwarfs with more extended star formation have lower central densities is more readily understood if the heating is the result of baryons \citep{2018arXiv180806634R}.

Given the current debate surrounding dwarf galaxies, it is unlikely that they will provide definitive answers to the nature of DM soon. However, the rate of DM self-interactions would scale with the local DM density, and (for the simplest models) with the local velocity-dispersion of DM particles. This means that more massive systems, such as galaxy clusters, hold promise as probes of the particle properties of DM.

For SIDM to explain the distribution of DM in dwarf and low-surface-brightness galaxy haloes requires a cross-section per unit mass $\sigma/m \gtrsim 0.5 \cmsg$ at DM--DM velocities of $30$--$100 \kms$ \citep[for a review see][]{2017arXiv170502358T}.\footnote{$\sigma(v_\mathrm{rel})$ is the DM--DM scattering cross-section at a relative velocity $v_\mathrm{rel}$, while $m$ is the mass of a DM particle.} Current constraints on the SIDM cross-section coming from cluster scales (which probe DM--DM velocities of $\sim 1000 \kms$ ), include $\sigma/m < 0.1 \cmsg$ \citep[][strong lensing arc statistics]{2001MNRAS.325..435M},  $\sigma/m < 0.3 \cmsg$ \citep[][subhalo evaporation]{2001ApJ...561...61G}, $\sigma/m \lesssim 1 \cmsg$ \citep[][cluster ellipticities]{2013MNRAS.430..105P} and $\sigma/m < 0.47 \cmsg$ (\citealt{2015Sci...347.1462H}, DM-galaxy offsets in merging clusters, though see \citealt{2017arXiv170105877W}).

If constraints on $\sigma / m$ from the literature are taken at face value, the door has been closed on a velocity-independent cross-section that can solve the `small-scale problems' with CDM. While a velocity dependence is naturally achieved in SIDM models where DM particles scatter through a Yukawa-like potential \citep{2010PhRvD..81h3522B,2010PhLB..692...70I,2010PhRvL.104o1301F,2011PhRvL.106q1302L,2013PhRvD..87k5007T,2013PhRvL.110k1301T,2014PhRvD..89k5017B,2014JCAP...05..047K,2015PhLB..751..201K,2017MPLA...3250038M}, other models of SIDM such as a self-coupled scalar \citep{2000PhRvD..62d1302B,2001NuPhB.619..709B,2002PhRvL..88i1304M,2015PhRvL.115b1301H} have a cross-section that is necessarily velocity-independent. Even those models for which velocity dependence is possible will typically be velocity-independent in some region of parameter space.\footnote{For example, if DM scatters through a Yukawa potential, then if the mediator mass ($m_\phi$) is greater than $\sim$1\% of the DM particle mass ($m_\chi$), the scattering cross-section will be independent of velocity for all astrophysically relevant DM velocities.} Assessing the robustness of constraints that indicate $\sigma/m \lesssim 0.5 \cmsg$ is therefore of vital importance, because these constraints rule out a large fraction of SIDM parameter space (and all velocity-independent cross-sections) from containing a viable explanation of the behaviour of DM in dwarf and low-surface-brightness galaxies.

SIDM constraints from galaxy clusters have had a chequered past, with constraints often overstated because of faulty assumptions. As an example, early work by \citet{2002ApJ...564...60M} argued that inside the radius at which particles would each interact on average once per Hubble time, the DM halo should be spherical. Combining this with a mass model of the galaxy cluster MS 2137-23 (derived from strongly-lensed gravitational arcs), which required the mass distribution to be aspherical at radii of $70 \kpc$, a constraint of $\sigma/m \lesssim 0.02 \cmsg$ was obtained. However, a more detailed study that made use of DM-only simulations with SIDM \citep{2013MNRAS.430..105P} found that this limit was severely over-stated, with a more realistic limit from halo shapes being $\sigma/m \lesssim 1 \cmsg$. The main drivers of this weakened constraint were that simulations showed that one scattering event per particle is not enough to remove all triaxiality from a DM halo, as well as the fact that lensing depends on the projected density, such that lensing measurements near the centre of the halo receive a contribution from large scales that are not affected by DM interactions.

Even the results of SIDM-only $N$-body simulations may not be adequate for making comparisons with observations. \citet{2018MNRAS.476L..20R} showed that baryons can have a significant effect on the distribution of SIDM within galaxy clusters, presenting high-resolution simulations of two galaxy clusters \citep[from the \ceagle sample of][]{2017MNRAS.471.1088B, 2017MNRAS.470.4186B} with an SIDM cross-section of $1 \cmsg$ and \eagle galaxy formation physics \citep{2015MNRAS.446..521S,2015MNRAS.450.1937C}. The responses of these two systems to the inclusion of baryons were starkly different. One of the two SIDM+baryons haloes had a total density profile almost indistinguishable from the CDM+baryons equivalent, while in the other system a large constant density core was present with SIDM+baryons that was virtually unchanged from the SIDM-only version of this system. These two different responses to including baryons into simulated SIDM clusters provided the motivation for performing simulations of large cosmological boxes with SIDM+baryons presented here. As well as addressing whether SIDM clusters do exhibit an enhanced diversity over their CDM counterparts (or whether one or both of the \citet{2018MNRAS.476L..20R} SIDM+baryons clusters were outliers), the presence of stars and gas in these simulations enables the generation of realistic mock data sets that can be used going forward to test existing \citep[e.g.][]{2001ApJ...561...61G, 2001MNRAS.325..435M, Randall:2008hs, 2015Sci...347.1462H, 2017MNRAS.464.3991H, 2017MNRAS.468.5004T, 2017MNRAS.469.1414K} 
or future methods of constraining the SIDM cross-section.

The paper is organised as follows. In Section~\ref{sect:sims} we describe the simulations, including the different SIDM models simulated. In Section~\ref{sect:density_profiles} we show the density profiles of our simulated clusters, showing their shapes in Section~\ref{sect:halo_shapes}. In Section~\ref{sect:lensing} we describe our method for producing strong lensing maps from our simulations, and then compare the strong lensing properties of our simulated clusters with an observed sample. Finally, we summarise our results in Section~\ref{sect:conclusions}.


\section{Numerical Simulations}
\label{sect:sims}

Our simulations combine the smoothed-particle hydrodynamics galaxy formation code used for \bahamas \citep{2017MNRAS.465.2936M} with the SIDM code that was introduced in \citet{2017MNRAS.465..569R}, with a more detailed description of the code available in \citet{myphd}. We refer the reader to these papers for more details, but outline the most relevant information below.

\subsection{Galaxy formation physics}
\label{sect:BAHAMAS_physics}

The \bahamas project \citep{2017MNRAS.465.2936M,McCarthy2018} consists of a suite of simulations designed to test the impact of baryonic physics on the interpretation of large-scale structure tests of cosmology. The majority of the simulations are of periodic boxes, $400 \, h^{-1} \, \mpc$ on a side, with $2 \times 1024^3$ particles. While the \bahamas simulations have been run with differing cosmologies, here we use only the \textit{WMAP} 9-yr cosmology\footnote{With $\Omega_\mathrm{m}=0.2793$, $\Omega_\mathrm{b}=0.0463$, $\Omega_\mathrm{\Lambda}=0.7207$, $\sigma_8 = 0.812$, $n_\mathrm{s} = 0.972$ and $h = 0.700$.} \citep{2013ApJS..208...19H} simulations, which have DM and (initial) baryon particle masses of $\num{5.5e9} \msun$ and $\num{1.1e9} \msun$, respectively. The Plummer-equivalent gravitational softening length is $5.7 \kpc$ in physical coordinates below $z=3$ and is fixed in comoving coordinates at higher redshifts.

\bahamas was run using a modified version of the {\sc Gadget-3} code \citep{Springel2005}.  The simulations include subgrid treatments for metal-dependent radiative cooling \citep{Wiersma2009a}, star formation \citep{Schaye2008}, stellar evolution and chemodynamics \citep{Wiersma2009b}, and stellar and AGN feedback \citep{DallaVecchia2008,Booth2009}, developed as part of the OWLS project (see \citealt{Schaye2010} and references therein).  For \bahamas, the parameters of the stellar and AGN feedback were adjusted so as to reproduce the observed present-day galaxy stellar mass function and the hot gas mass within groups and clusters of galaxies.  \bahamas also reproduces a large number of observables including the richness, size, and stellar mass functions of galaxy groups, the dynamics of satellite galaxies as a function of halo mass, the local stellar mass autocorrelation function, and the stacked weak lensing and thermal Sunyaev-Zel'dovich (SZ) signals of systems binned by stellar mass \citep{2017MNRAS.465.2936M,2018MNRAS.480.3338J} suggesting that it employs a realistic model of galaxy formation.

We found that the galaxy stellar mass function, gas fractions and X-ray luminosities of clusters, to which the subgrid model parameters were calibrated, remain virtually unchanged with the inclusion of SIDM, such that re-calibration is unnecessary. Nevertheless, some of the properties discussed in this paper may be sensitive to our adopted model of baryonic physics.  We explore this question further in Appendix~\ref{App:subgrid}, using the \bahamas variation models (``low AGN'' and ``hi AGN'') from \citet{McCarthy2018}.

\subsection{SIDM models}

As well as CDM, we simulated three different SIDM models. Two of the models have velocity-independent cross-sections and isotropic scattering, with $\sigma/m = 0.1$ and $1 \cmsg$; SIDM0.1 and SIDM1 respectively. The final cross-section (labelled vdSIDM) is velocity-dependent, and corresponds to DM particles scattering though a Yukawa potential. This model is described by 3 parameters: the DM mass $m_\chi$, the mediator mass $m_\phi$, and a coupling strength $\alpha_\chi$. In the Born limit, $\alpha_\chi m_\chi \ll m_\phi$, the differential cross-section is \citep{2010PhLB..692...70I,2013PhRvD..87k5007T}
\begin{equation}
\label{eq:yukawa_differential_cross-sect}
\frac{\D \sigma}{\D \Omega} = \frac{\alpha_\chi^2}{m_\chi^2 \left(m_\phi^2/m_\chi^2  + v^2 \sin^2 \frac{\theta}{2} \right)^2 }, 
\end{equation}
where $v$ is the relative velocity between two DM particles, and $\theta$ the polar scattering angle in the centre of mass frame of the two particles. This can be written as \citep{2017MNRAS.467.4719R}
\begin{equation}
\label{eq:yukawa_differential_cross-sect_alt}
\frac{\D \sigma}{\D \Omega} = \frac{\sigma_0}{4 \pi (1 + \frac{v^2}{w^2} \sin^2 \frac{\theta}{2})^2 },
\end{equation}
where $w= m_\phi c / m_\chi$ is a characteristic velocity below which the scattering is roughly isotropic with $\sigma \approx \sigma_0$, and above which the cross-section decreases with increasing velocity, also becoming more anisotropic, favouring scattering by small angles. Our velocity-dependent model has  $m_\chi = 0.15 \gev$, $m_\phi = 0.28 \kev$ and $\alpha_\chi = 6.74 \times 10^{-6}$, corresponding to $\sigma_0 = 3.04 \cmsg$, $w = 560 \kms$. These model parameters were chosen to roughly reproduce the behaviour as a function of velocity of the best-fitting cross-section in \citet{2016PhRvL.116d1302K}, which was shown to successfully explain the density profiles of systems ranging from dwarf galaxies to galaxy clusters, though the $\approx 3 \cmsg$ cross-section at low velocities may be in tension with a recent analysis of stellar kinematics in the Draco dwarf galaxy \citep{2018MNRAS.481..860R}.

\subsection{SIDM implementation}

Our method of simulating SIDM is that described in \citet{2017MNRAS.465..569R}, which uses a similar Monte-Carlo approach to implement DM scattering as other recent SIDM simulations \citep{2012MNRAS.423.3740V,2013MNRAS.430...81R, 2013MNRAS.430..105P, 2013MNRAS.430.1722V, 2013MNRAS.431L..20Z, 2014MNRAS.444.3684V, 2015MNRAS.453...29E, 2017MNRAS.468.2283C, 2017MNRAS.469.1414K, 2017MNRAS.469.2845D, 2018MNRAS.474..746B, 2018ApJ...853..109E, 2018MNRAS.479..359S}. At each time-step, particles search locally for neighbours, with random numbers drawn to see which nearby pairs scatter. The probability for a pair of particles to scatter depends on their relative velocity and the cross-section for scattering, which itself can be a function of the relative velocity. The search region around each particle is a sphere, with a radius equal to the gravitational softening length. For vdSIDM we do not follow the majority of previous work \citep[e.g.][]{2012MNRAS.423.3740V,2013MNRAS.430.1722V, 2013MNRAS.431L..20Z, 2014MNRAS.444.3684V, 2016MNRAS.460.1399V}, where anisotropic cross-sections were simulated using isotropic scattering and an effective cross-section, instead implementing the differential cross-section from equation~\eqref{eq:yukawa_differential_cross-sect_alt} directly, using the method described in \citet{2017MNRAS.467.4719R}.

\begin{figure}
        \centering
        \includegraphics[width=\columnwidth]{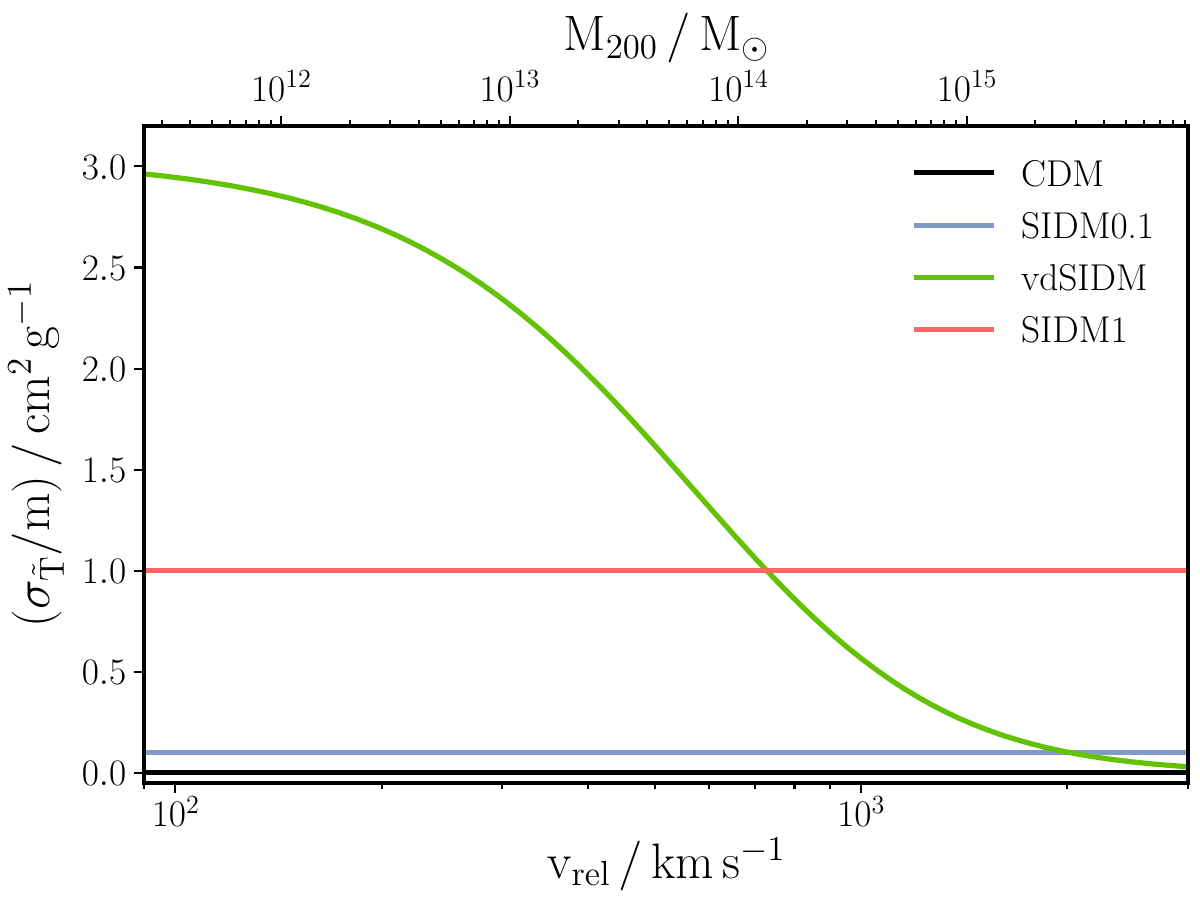}
	\caption{The momentum transfer cross-section for our four simulated DM models, as a function of the relative velocity between DM particles. This relative velocity has been roughly converted to a $z=0$ halo mass (along the top of the figure) using $v_\mathrm{rel} = \sqrt{G \, M_{200} / r_{200}}$. The colour scheme used here for different DM models is continued throughout the rest of the paper.}
	\label{fig:cross-sections}
\end{figure}

In Fig.~\ref{fig:cross-sections} we plot the cross-section as a function of relative velocity for our four simulated DM models. Specifically, we plot the momentum transfer cross-section
\begin{equation}
\label{eq:sigT}
\sigma_{\tilde{T}} \equiv 2 \int (1 - |\cos \theta|) \frac{\D \sigma}{\D \Omega} \D \Omega ,
\end{equation}
which has been shown to be a more relevant quantity than the total cross-section for determining the rate at which cores form in isolated DM haloes \citep{2017MNRAS.467.4719R}.\footnote{This differs by a factor of two from the definition in \citet{2017MNRAS.467.4719R}, but has been chosen such that for isotropic scattering, $\sigma = \sigma_{\tilde{T}}$.} The $1 - \cos \theta$ term comes from weighting scatterings by the amount of momentum they transfer along the collision axis \citep{2014MNRAS.437.2865K}, with $|\cos \theta|$ being used because if particles scatter by angles greater than $90^{\circ}$, then the particles could be re-labelled\footnote{As an illustrative example: if two indistinguishable particles scatter by $180^{\circ}$, then while a large amount of momentum is transferred, the result is the same as if the two particles had not interacted at all.} such that the scattering was by less than $90^{\circ}$.

When considering a velocity-dependent cross-section, it would be useful if for a given system this could be mapped onto the velocity-independent and isotropic cross-section that would produce the most similar DM distribution. One might assume that this cross-section would be approximately equal to $\sigma_{\tilde{T}}(|v_\mathrm{rel}|)$, where $|v_\mathrm{rel}|$ is the mean pairwise velocity of particles within the halo. Even the definition of $|v_\mathrm{rel}|$ has subtleties (do particle pairs receive equal weight, or should it be weighted by the scattering probability of those pairs), and the complicated assembly history of haloes means that a halo with some velocity dispersion now, likely had a lower velocity dispersion in the past. Nevertheless, in Fig.~\ref{fig:cross-sections} we crudely relate the pairwise velocity of particles to a halo mass, using $v_\mathrm{rel} = \sqrt{G \, M_{200} / r_{200}}$ (at $z=0$).\footnote{We define $r_{200}$ as the radius at which the mean enclosed density is 200 times the critical density, and $M_{200}$ as the mass within $r_{200}$.} From Fig.~\ref{fig:cross-sections} we can expect haloes with $M_\mathrm{200} \sim 10^{14} \msun$ to look similar with vdSIDM and SIDM1, while in more massive haloes vdSIDM will behave more like SIDM0.1.
 

\section{Density profiles}
\label{sect:density_profiles}

The main motivation for SIDM has been the effect that self-interactions have on DM density profiles, especially for dwarf galaxies. The primary effect is a reduction in the density of DM in the central regions of a halo when compared with a CDM equivalent, though in the presence of a dense baryonic component this effect can be reversed \citep{2018MNRAS.479..359S} by decreasing the timescale on which haloes undergo \emph{core-collapse}.\footnote{Also referred to as `gravothermal collapse', or the `gravothermal catastrophe' and studied in the context of SIDM by \citet{2002ApJ...568..475B} and \citet{2011MNRAS.415.1125K} amongst others.} This same effect is expected in galaxy clusters, where the larger velocity dispersions lead to higher rates of scattering than in dwarf galaxies, at least for a velocity-independent cross-section, making SIDM-induced galaxy cluster cores more prominent than their dwarf galaxy counterparts. 

\begin{figure*}
        \centering
        \includegraphics[width=\textwidth]{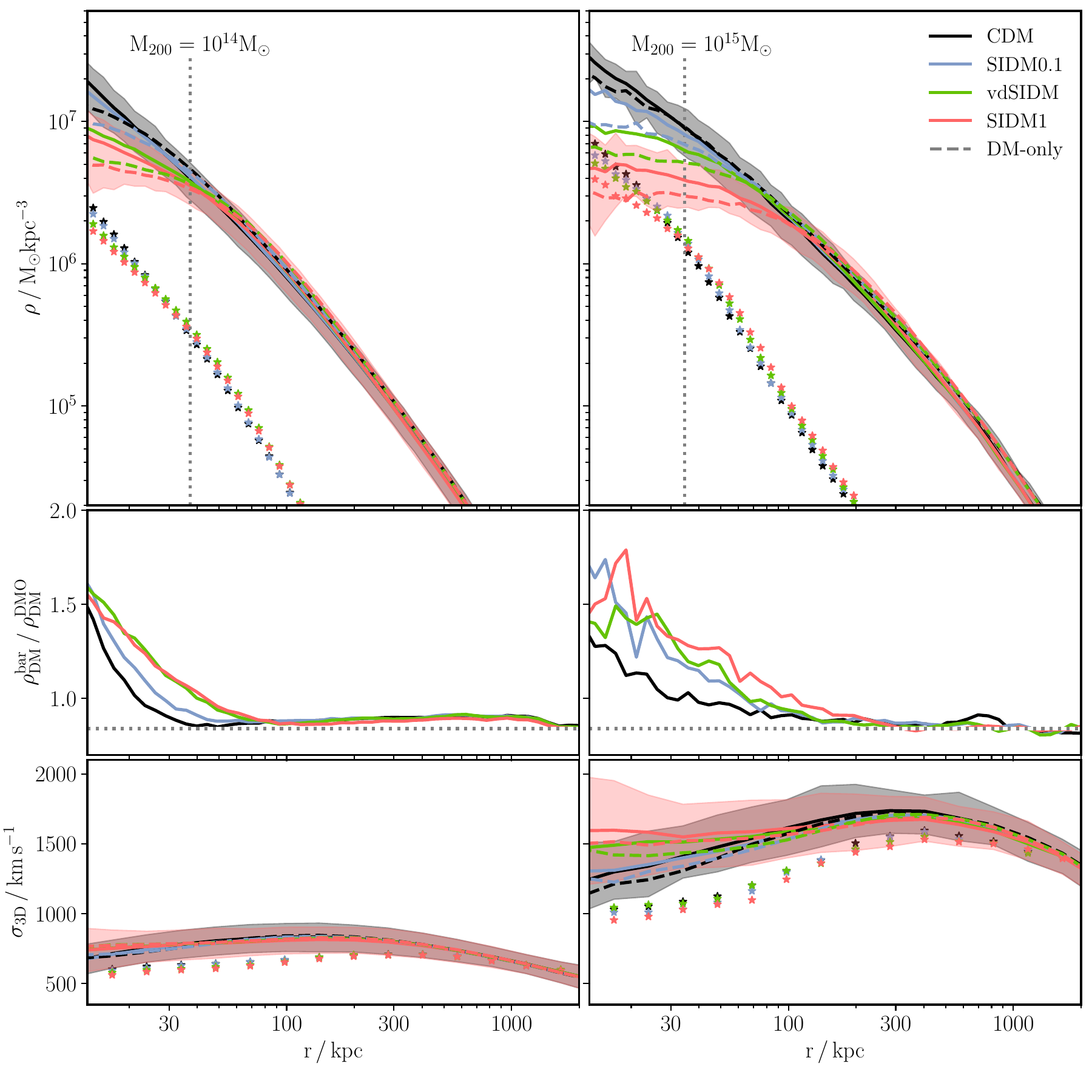}
	\caption{Top: stacked radial density profiles of our simulated clusters at $z=0$, with the left and right panels showing the results for clusters with $M_{200} \approx 10^{14}$ and $10^{15} \msun$ respectively. The solid lines show the mean density profile of the DM from the full physics runs, while the dashed lines show the DM-only equivalents. The stars show the stellar density profile from the full physics runs. The black and red shaded regions show the 16-84th percentile ranges of DM profiles from the CDM and SIDM1 full physics simulations respectively. The vertical dotted lines show the median convergence radii (equation~\ref{P03_conv}) calculated from the DM in the full physics CDM simulation. Middle: the mean DM density from simulations with full physics divided by the mean DM density from the corresponding DM-only simulations. The dotted lines show $\Omega_\mathrm{DM} / \Omega_\mathrm{m}$ for our assumed cosmology. Bottom: mean velocity dispersion profiles, with the lines and stars referring to the same components as in the top panels.}
	\label{fig:density_profiles}
\end{figure*}

\subsection{Method}

In Fig.~\ref{fig:density_profiles} we plot the $z=0$ spherically-averaged density profiles of stars and DM from our simulations with different DM models and \bahamas physics, as well as the density profiles from DM-only simulations. The centre of the halo is defined by the location of the most-bound particle, and the density is calculated from the summed mass of particles within logarithmically-spaced spherical shells. The density profiles shown are the mean density as a function of radius for haloes within two fairly narrow mass bins centred on $M_{200} = 10^{14}$ and $10^{15} \msun$. The virial radii of $10^{14}$ and $10^{15} \msun$ haloes at $z=0$ are $r_{200} = 0.96$ and $2.08 \mpc$ respectively. The left panel corresponds to haloes with $13.9 < \log_{10} \left[ M_{200} / M_\odot \right] < 14.1$ and the right panel to $14.8 < \log_{10} \left[ M_{200} / M_\odot \right] < 15.2$, with these bins containing approximately 1000 and 40 haloes respectively. To indicate the typical size of the scatter about these mean density profiles, the 16-84th percentile regions from the CDM+baryons and SIDM1+baryons simulations have also been shown.

Determining the smallest radius at which we should trust our simulated density profiles is a difficult task. In the case of DM-only simulations, this radius can be determined by running simulations at different resolutions and observing the radial range over which their density profiles agree. This task is considerably more difficult for simulations that include subgrid models of galaxy formation physics, as physical properties of the feedback (such as the energy injected per event or their frequency) depend on the simulation's resolution.\footnote{For a detailed discussion of convergence in hydrodynamical simulations of galaxy formation see \citet{2015MNRAS.446..521S}.} Here, we follow \citet{2015MNRAS.451.1247S} and define the convergence radius, $r_\mathrm{conv}$, as the smallest radius for which 
\begin{equation}
0.33 \leq \frac{\sqrt{200}}{8} \sqrt{\frac{4 \pi \rho_\mathrm{crit}}{3 \, m_\mathrm{DM}}} \frac{\sqrt{N(<r_\mathrm{conv})}}{\ln N(<r_\mathrm{conv})} r_\mathrm{conv}^{3/2},
\label{P03_conv}
\end{equation}
where $\rho_\mathrm{crit}$ is the critical density, $m_\mathrm{DM}$ the mass of the simulation DM particles and $N(<r)$ the number of DM particles within a radius $r$. This criterion relates to the two-body relaxation timescale of particles and is inspired by the DM-only convergence studies from \citet{2003MNRAS.338...14P}. We calculated $r_\mathrm{conv}$ for all of our simulated haloes, and show the median $r_\mathrm{conv}$ from the CDM full physics haloes (in the respective halo mass bins) in Fig.~\ref{fig:density_profiles}. Note that our convergence radii depend on the distribution and properties of only the DM particles in our simulations, and that given the lower densities with SIDM the convergence radius as defined in equation~\eqref{P03_conv} is formally larger with SIDM than CDM. A thorough study of convergence of simulated SIDM density profiles has not been performed, but SIDM-only convergence tests typically indicate that SIDM density profiles are better converged than their CDM counterparts \citep{2012MNRAS.423.3740V, myphd}. This is not surprising given that gravitational two body interactions that artificially soften simulated central CDM cusps, are relatively unimportant when simulating a model with physical two body interactions, as is the case with SIDM.

To compare the relative effect of including baryons with the different DM models, we plot the ratios of the DM densities from our simulations with \bahamas physics to the DM densities from the corresponding DM-only simulations in the middle panels of Fig.~\ref{fig:density_profiles}. Because all of the matter in a DM-only simulation is modelled as DM, the DM density in a simulated DM-only universe is larger than in the corresponding DM+baryons universe by a factor of $\Omega_\mathrm{m} / \Omega_\mathrm{DM}$. We therefore plot a horizontal line at $\Omega_\mathrm{DM} / \Omega_\mathrm{m} \approx 0.84$.

Finally, in the bottom panels of Fig.~\ref{fig:density_profiles} we plot the 3D velocity dispersion profiles of both DM and star particles in our simulations. These are calculated by taking all $N$ particles within logarithmically-spaced spherical shells (wider than in the case of the density profiles) and then using
\begin{equation}
\sigma_\mathrm{3D}^2 = \frac{1}{N} \sum_{i=1}^N |\vect{v}_i - \vect{v}_\mathrm{CoM}|^2,
\label{sigma_3D}
\end{equation}
where $\vect{v}_\mathrm{CoM}$ is the mass-weighted mean velocity of all particles in the halo. As in the top panels, we show both the DM-only and \bahamas physics results, and the 16-84th percentile regions for CDM+baryons and SIDM1+baryons.

\begin{figure}
        \centering
        \includegraphics[width=\columnwidth]{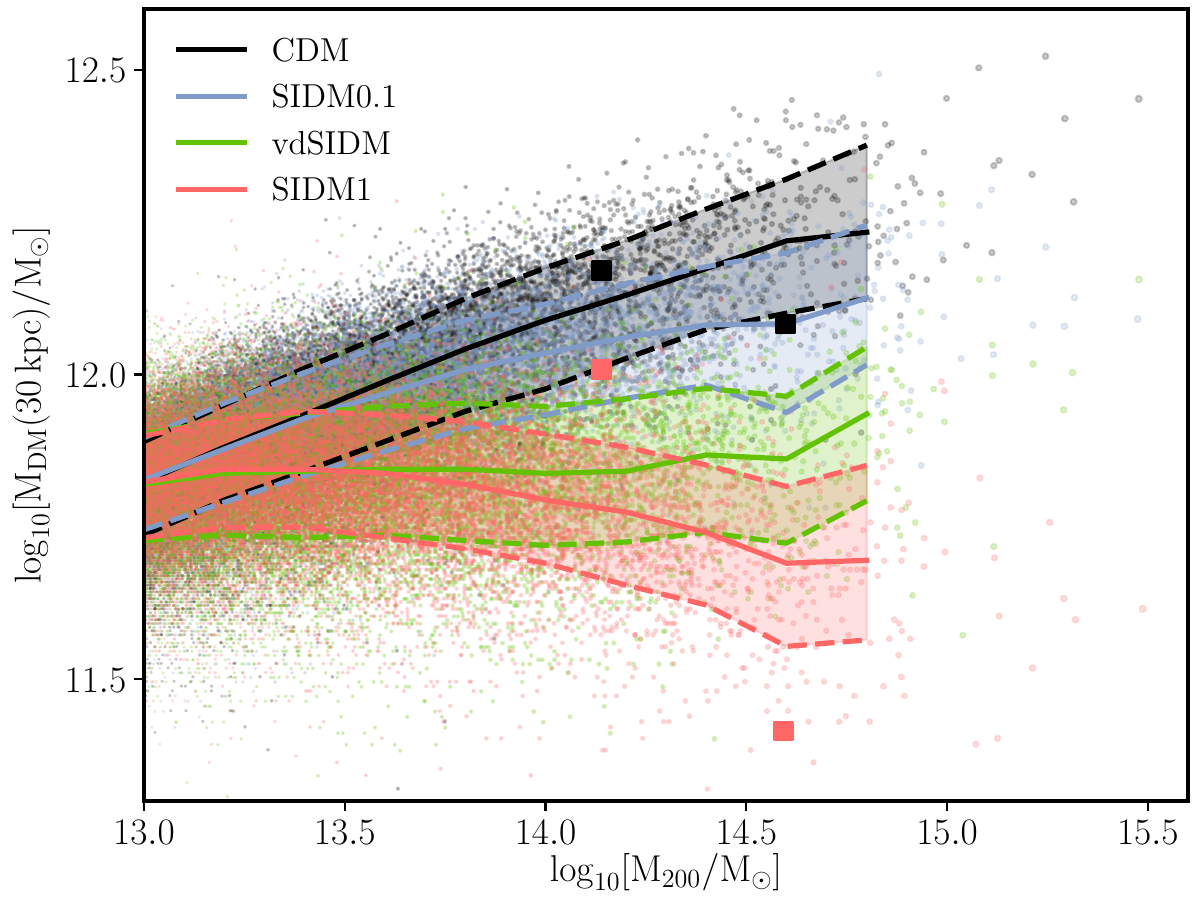}
	\caption{The DM mass within a $30 \kpc$ spherical aperture as a function of halo mass, for haloes at $z=0$. Points show individual haloes, while the solid lines show the median relations (measured in 0.2 dex $M_{200}$ bins), with shaded regions denoting the 16-84th percentile ranges. The squares show where the two \ceagle clusters simulated with CDM and SIDM1 presented in \citet{2018MNRAS.476L..20R} lie in this plot.}
	\label{fig:M200_Mdm30}
\end{figure}

\begin{figure}
        \centering
        \includegraphics[width=\columnwidth]{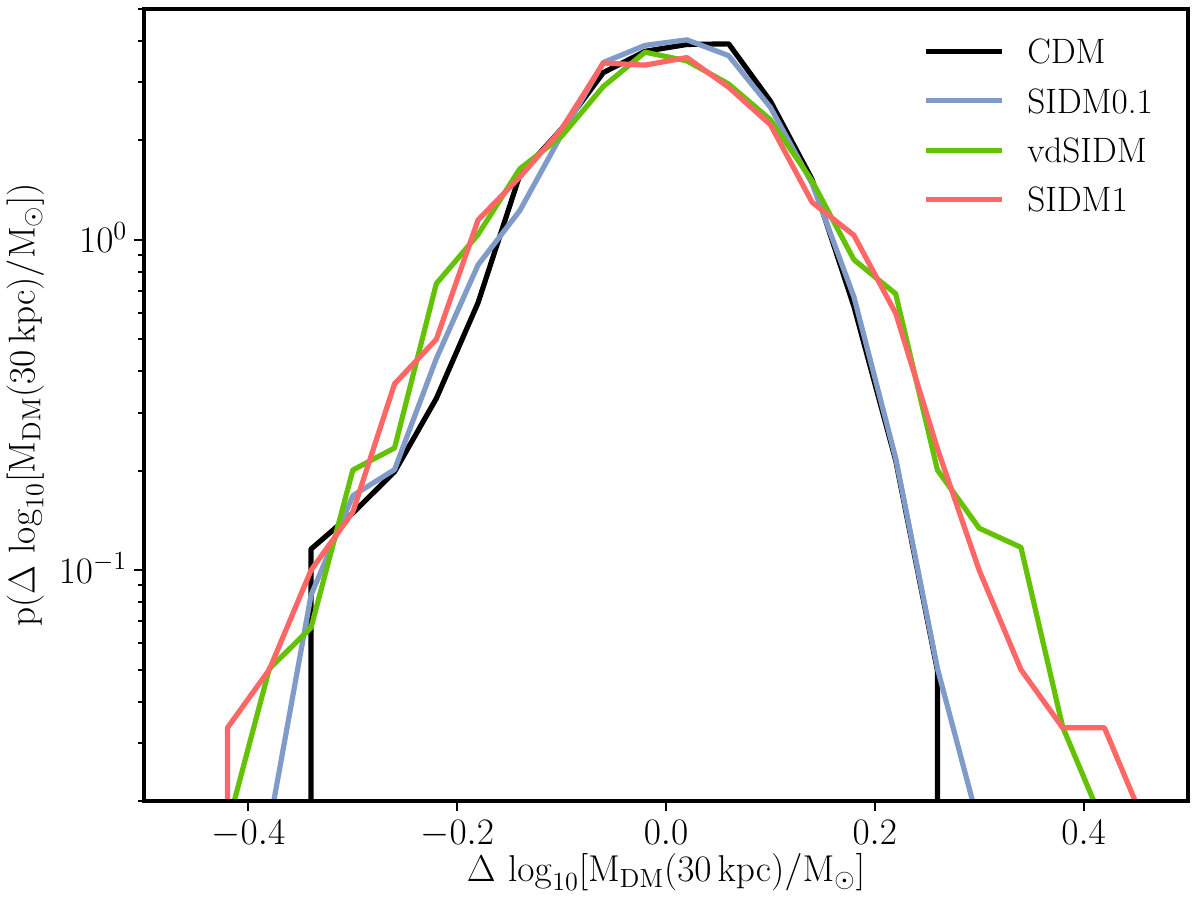}
	\caption{The PDF of $\Delta \log_{10} \left[ M_\mathrm{DM} (30 \kpc) / \msun \right]$, for all haloes with $14.0 < \log_{10} \left[ M_{200} / \msun \right] < 14.8$. $\Delta \log_{10} \left[ M_\mathrm{DM} (30 \kpc) / \msun \right]$ is defined for each halo as the difference in $ \log_{10} \left[ M_\mathrm{DM} (30 \kpc) / \msun \right]$ between that halo and the corresponding median line shown in Fig.~\ref{fig:M200_Mdm30}.}
	\label{fig:delta_Mdm30}
\end{figure}

\subsection{Results}

DM interactions decrease the central density of haloes, with larger cross-sections leading to greater decreases in density (see the top panels of Fig.~\ref{fig:density_profiles}). 
For the vdSIDM model, we find 
$M_{200} \sim 10^{14} \msun$ haloes with density profiles similar to those with SIDM1, while for higher halo masses, the behaviour is in between that of SIDM1 and SIDM0.1. This is expected from Fig.~\ref{fig:cross-sections}, suggesting that density-profile predictions for a velocity-dependent cross-section can be made using an appropriately-matched velocity-independent one, as has been assumed in previous work \citep[e.g.][]{2016PhRvL.116d1302K}.

For all DM models, we find that baryons increase the density of DM in the centres of haloes, compared with DM-only runs. However, this increase occurs over larger scales in SIDM models, out to $\sim5\%$ of $r_{200}$ for SIDM1 -- compared with half that with CDM (see the middle panels of Fig.~\ref{fig:density_profiles}). 

The scatter in the logarithm of the central densities is also slightly larger in SIDM1 than CDM, but this can at least partly be attributed to the fact that the lower density of SIDM halo centres leads to fewer particles per radial bin and so increased Poisson noise.\footnote{Given the size of our radial bins, for $M_{200} = 10^{14} \msun$ at $r=12 \kpc$ the density at the 16th percentile of the SIDM1 range ($\sim \num{3e6} \msun \kpc^{-3}$) only corresponds to 7 DM particles.} To investigate the extent of any enhanced diversity of SIDM clusters, we plot the DM mass within a $30 \kpc$ spherical aperture as a function of halo mass in Fig.~\ref{fig:M200_Mdm30}. While the trends for the different DM models are starkly different, the spread in $\log \left[ M_\mathrm{DM}(30 \kpc) / M_\odot \right]$ about the corresponding median relationship is similar between the different DM models. We show this in Fig.~\ref{fig:delta_Mdm30} where we plot the probability density function of $\Delta \log_{10} \left[ M_\mathrm{DM} (30 \kpc) / \msun \right]$, which is the difference between the value of $\log_{10} \left[ M_\mathrm{DM} (30 \kpc) / \msun \right]$ and the median lines in Fig.~\ref{fig:M200_Mdm30}. With vdSIDM and SIDM1 there is a slight tail towards high values, evident also in Fig.~\ref{fig:M200_Mdm30} where, for example, at the high mass end ($M_{200} > 10^{14.5} \msun$) there are red points scattered up into the black shaded region. 
While there is a small enhancement in the scatter of central densities with SIDM, a key prediction from these simulations is that this scatter is not substantial, so a moderate number of well studied clusters may suffice to place robust limits on the SIDM cross-section.


\subsection{Discussion}
\label{sect:dens_discuss}

The two clusters simulated with SIDM+baryons by \citet{2018MNRAS.476L..20R} as part of the \ceagle project \citep{2017MNRAS.471.1088B, 2017MNRAS.470.4186B} had starkly different central density profiles. To understand how these simulated clusters fit in with those presented here, we plot both the CDM and SIDM1 versions of the two \ceagle clusters in Fig.~\ref{fig:M200_Mdm30}. While the \ceagle-SIDM1 haloes lie within the locus of \bahamas-SIDM1 clusters, they are $\approx 2 \sigma$ outliers in the context of the \bahamas-SIDM1 simulations. Meanwhile the \ceagle-CDM clusters are more typical of the population of \bahamas-CDM clusters.

It may simply be a coincidence that the two \ceagle-SIDM1 clusters ended up being outliers (in different directions) in terms of their central densities, either due to their particular assembly histories, or as a result of the chaotic-like behaviour of cosmological simulations \citep{2018arXiv180305445K, 2018arXiv180707084G}. Alternatively, the hint of increased diversity with SIDM seen in \ceagle, but not \bahamas, may depend sensitively on details of the simulations. In particular, the mechanism to produce centrally-dense SIDM clusters described in \citet{2018MNRAS.476L..20R} relied on stars dominating the gravitational potential on sufficiently small scales.
The stellar masses of central galaxies in \ceagle are 0.3 to 0.6 dex above their observed counterparts \citep[see the right-hand panel Fig.~4 of][]{2017MNRAS.470.4186B}, so the influence of baryons on the SIDM distribution at the centre of galaxy clusters is likely overestimated in \ceagle. The stellar masses of \bahamas central galaxies are lower than those in \ceagle, in better agreement with observed systems \citep{2017MNRAS.465.2936M}. 
The effect of baryons in our \bahamas-SIDM simulations may therefore be more realistic than in \ceagle-SIDM.

Another difference between \bahamas and \ceagle is resolution, with the scales on which stars dominate the potential ($< 10 \kpc$) being not well-resolved in \bahamas. This may be suppressing the impact of baryons on our SIDM density profiles, though whether this is happening will be hard to address without a larger number of clusters simulated at higher resolution with SIDM+baryons.

Aside from changes to the DM density, the different DM models also lead to different stellar density profiles in the inner regions. The starkest example is for the $10^{15} \msun$ haloes, where SIDM1 produces stellar density profiles which flatten in the centre, with a density at $10 \kpc$ similar to that of the DM. While this could potentially be used to constrain the SIDM cross-section, the stellar density profiles of simulated clusters are sensitive to both resolution and the details of AGN feedback \citep[e.g.][]{2011MNRAS.414..195T}. In Appendix~\ref{App:subgrid} we show how changing the temperature to which gas is heated by AGN alters the stellar and DM density profiles (with CDM). We find that the DM density profiles are relatively unaffected, but that the stellar density profiles change substantially, confirming that the stellar density profiles are not robustly predicted from our simulations. 

While the DM density profile can in principle be inferred from observations, these inferences are fraught with difficulty. For example, \citet{2013ApJ...765...25N} inferred the DM density profiles of clusters using a combination of strong and weak lensing as well as stellar kinematics. They found that the DM density profiles in the inner $30 \kpc$ were significantly shallower than the Navarro-Frenk-White (NFW) profile \citep{1997ApJ...490..493N} predicted with CDM. \citet{2015MNRAS.452..343S} presented a set of simulated CDM clusters with total density profiles similar to those inferred by \citet{2013ApJ...765...25N}, and also with similar surface brightness and line-of-sight velocity profiles for the central galaxies. However, the \citet{2015MNRAS.452..343S} clusters had DM density profiles that followed the NFW prediction. \citet{2015MNRAS.452..343S} suggested that this discrepancy could result from \citet{2013ApJ...765...25N} incorrectly assuming stellar orbits to be isotropic, or from using an incorrect stellar mass-to-light ratio. Whatever the reason for the discrepancy, this case is a good example of why comparisons between observations and simulations are often best-done in the observed quantities, i.e. \emph{forward modelling}, rather than inferring physical quantities from the observations. We provide an example of this in Section~\ref{sect:lensing} where we calculate the strong-lensing properties of our simulated clusters. Another example is in \citet{2019MNRAS.488.1572H}, where they looked at the offsets between peaks in the projected stellar and DM distributions of the simulations we present here.


\section{Halo shapes}
\label{sect:halo_shapes}

Aside from re-distributing energy between DM particles, DM self-interactions change the directions of DM particle orbits, leading to more isotropic velocity distributions. This in turn leads to more spherical DM density distributions, though a system with an isotropic velocity distribution can still exhibit ellipticity in its density and resulting potential \citep{2017JCAP...05..022A}. As mentioned in the introduction, early analytical work on the sphericity of galaxy clusters suggested exceptionally tight constraints on the SIDM cross-section \citep{2002ApJ...564...60M}, which SIDM-only simulations have shown to be over-stated \citep{2000ApJ...544L..87Y, 2013MNRAS.430..105P}. Recently, \citet{2018MNRAS.474..746B} presented simulations of 28 SIDM-only galaxy clusters, and showed that halo shapes were affected by SIDM on larger scales than density profiles, which could make halo shapes a test of DM self-interactions that is less sensitive to the details of baryonic physics than density profiles. 

\subsection{Method}

Our shape definition uses the location of the most bound particle as the centre of the halo, with the positions of particles defined with respect to that point and all spherical/ellipsoidal search volumes centred there also. To calculate the shape of a halo within a radius $r$, we begin by finding all particles in a sphere of radius $r$. The \emph{reduced inertia tensor}\footnote{We call this an `inertia tensor' for consistency with previous work, though it is not the tensor relating angular velocities to angular momenta.}
\begin{equation}
\tilde{I}_{ij} \equiv \sum_n \frac{x_{i,n} \, x_{j,n} \, m_{n}}{r_n^2} \, \Big/  \, \sum_n m_{n}
\label{mass_tensor}
\end{equation}
is calculated for this distribution of particles, where $(x_{1,n},x_{2,n},x_{3,n})$ are the coordinates of the $n$th particle, which has mass $m_{n}$. Initially $r_n$ is just the distance of the $n$th particle from the centre of the halo: $r_n = \sqrt{x_{1,n}^2 + x_{2,n}^2 + x_{3,n}^2}$. We label the eigenvalues of $\tilde{I}_{ij}$ as $a^2$, $b^2$ and $c^2$, with corresponding eigenvectors $\vect{e_1}$, $\vect{e_2}$ and  $\vect{e_3}$, and with $a \geq b \geq c$. The axis ratios are defined by $s = c/a$ and $q=b/a$.

Our process is iterative, stopping when subsequent iterations agree on both axis-ratios ($q$ and $s$) to better than 1\%. Specifically, we stop the iteration when 
\begin{equation}
\left( \frac{q_{i} - q_{i \text{-} 1}}{q_{i \text{-} 1}} \right)^2 + \left( \frac{s_{i} - s_{i \text{-} 1}}{s_{i \text{-} 1}} \right)^2 \leq \epsilon_\mathrm{conv}^2
\label{halo_shape_convergence}
\end{equation}
where the subscripts $i$ and $i - 1$ refer to the current values and the values from the previous iteration respectively, and with $\epsilon_\mathrm{conv} = 0.01$. Each iteration uses the eigenvectors from the previous iteration as a coordinate basis (i.e. for a particle at $\vect{x_n}$: $x_{1,n,i} = \vect{x_n} \cdot \vect{e_{1,i \text{-} 1}}$). The procedure from the first step is repeated, but defining $r_n$ as an \emph{ellipsoidal radius}
\begin{equation}
r_n = \sqrt{x_{1,n}^2 + x_{2,n}^2/q^2 + x_{3,n}^2/s^2} \, ,
\label{elliptical_radius}
\end{equation}
and with the sum in equation~\eqref{mass_tensor} over all particles with $r_n < (q\,s)^{-1/3} r$. Note that this corresponds to keeping the volume within which particles contribute to $\tilde{I}_{ij}$ fixed, while summing over particles with $r_n < r$ would keep the semi-major axis of the ellipsoid within which particles contribute to $\tilde{I}_{ij}$ equal to the radius of the initial sphere. We found that this distinction made little difference to our qualitative findings \citep[as also found by][]{2013MNRAS.430..105P}, and we use the $r_n < (q\,s)^{-1/3} r$ definition throughout this work.

\begin{figure*}
        \centering
        \includegraphics[width=\textwidth]{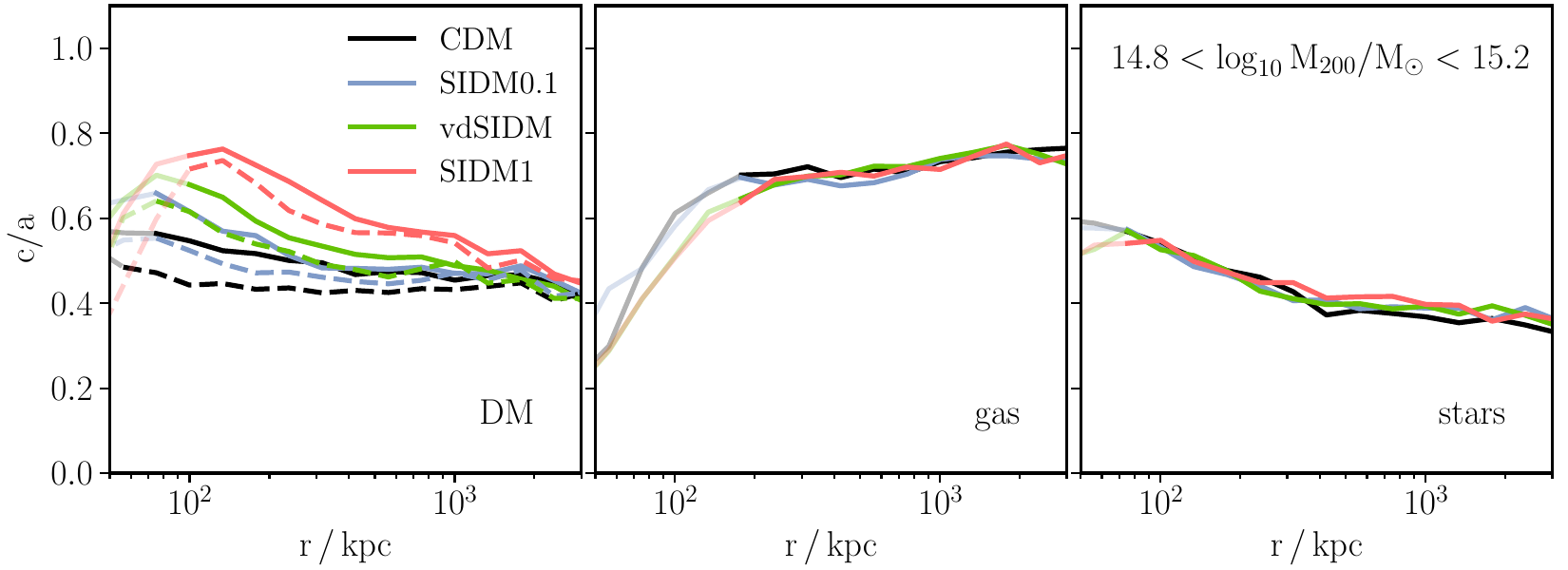}
        \includegraphics[width=\textwidth]{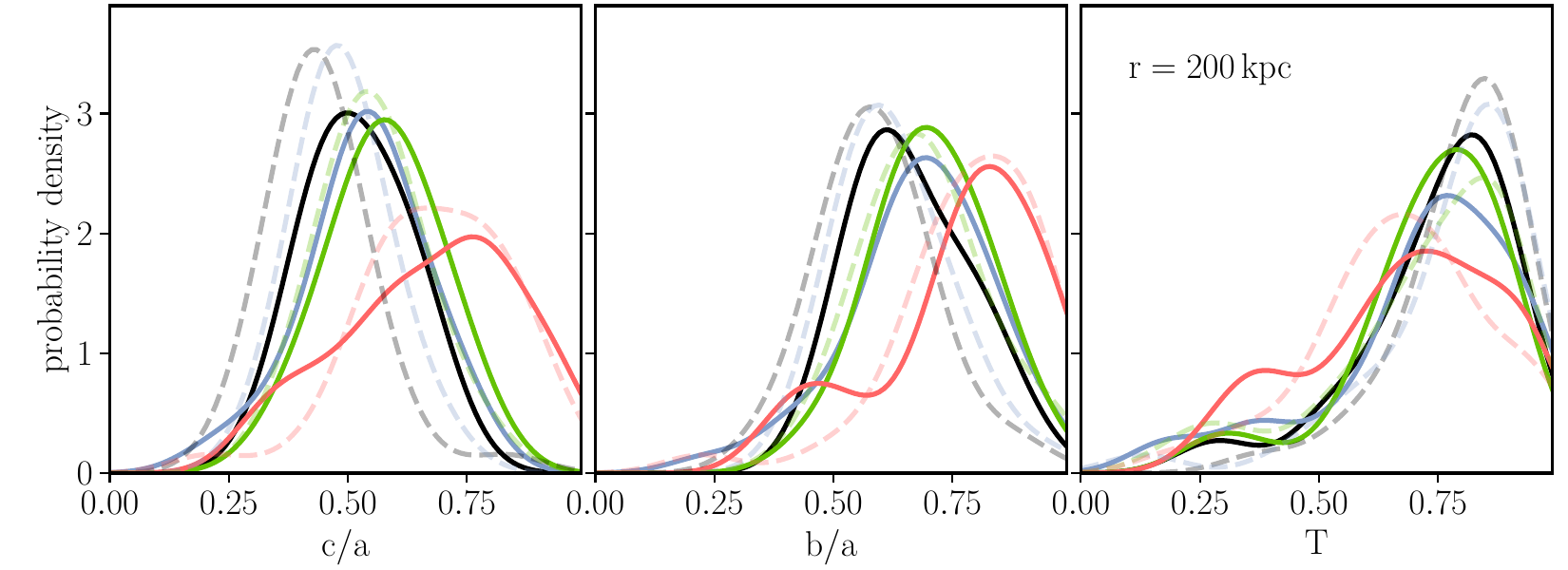}
	\caption{Top: the minor to major axis ratios for the DM, gas and stars from our different simulations at $z=0$, with the DM-only results represented by dashed lines. The lines shown are the medians from the $\sim 40$ haloes (per simulation) in the mass range $10^{14.8}$--$10^{15.2} \msun$, and become faded at radii where there are an insufficient number of particles to trust the shape measurements. Our definition of halo shape and the method used to calculate it is described in Section~\ref{sect:halo_shapes}. While the DM halo becomes rounder (especially in the inner regions) with increasing cross-section, this is not obviously reflected in the shapes of either the gas or stars. Bottom: the distributions of DM axis ratios at $r=200 \kpc$, for the same haloes in the top panel. The triaxiality is defined by $T=(a^2 - b^2)/(a^2 - c^2)$, with values $T \lesssim 1/3$ and $T \gtrsim 2/3$ representing oblate and prolate distributions respectively. The DM-only lines in the bottom panels are faded to avoid distracting from the full physics results, not because we do not trust them.}
	\label{fig:halo_shapes_M15}
\end{figure*}

\subsection{Results}

In agreement with previous work \citep{2013MNRAS.430..105P, 2018MNRAS.474..746B}, we find that SIDM makes DM haloes rounder, especially in the inner regions of haloes where the scattering rates are highest. In Fig.~\ref{fig:halo_shapes_M15} we plot the median minor-to-major axis ratios as a function of radius, for haloes with $M_{200} \approx 10^{15} \msun$. The lines are semi-transparent at radii where fewer than 800 particles contribute to the shape measurement, because we found from tests described in Appendix~\ref{App:halo_shape_tests} that at least this number of particles was required for a robust determination of the halo shape. The trend with cross-section is similar to that for the density profiles in the same halo mass range, with the behaviour of vdSIDM being intermediate to SIDM0.1 and SIDM1. The size of the change in median axis ratio going from CDM to SIDM1 is significantly larger than the change from DM-only to including the effects of baryons \citep[for more on the effect of baryons on halo shapes see e.g.][]{2013MNRAS.429.3316B}.

The shape differences persist to radii beyond where there is a notable change in the density profiles, as shown for the case of SIDM-only clusters in \citet{2018MNRAS.474..746B}. While this is partly driven by our use of all enclosed mass in the reduced inertia tensor (so the round central regions contribute to the calculation of $c/a$ even in the outskirts), when redoing our analysis using ellipsoidal shells \citep[as done in][]{2018MNRAS.474..746B} there is still a clear trend in $c/a$ with cross-section out to $\sim 1 \mpc$.

In the bottom panels of Fig.~\ref{fig:halo_shapes_M15} we show distributions of the two axis ratios as well as the triaxiality parameter, $T=(a^2 - b^2)/(a^2 - c^2),$\footnote{Values of $T \lesssim 1/3$ and $T \gtrsim 2/3$ represent oblate and prolate distributions, respectively.} for the DM components of our high-mass haloes. These were measured at $r=200 \kpc$, for the same set of $\approx 40$ haloes used in the top panels. Due to the low number of haloes used, these distributions were smoothed using kernel density estimation.\footnote{We used \code{KernelDensity} from scikit-learn \citep{scikit-learn}, using a Gaussian kernel with a \emph{bandwidth} (i.e. standard deviation) of 0.08.} Both SIDM and baryons affect the 3D shapes in a similar manner, in the sense that they primarily increase $c/a$, making the haloes not just more spherical, but also less prolate. We stress that the DM-only lines in the bottom panels of Fig.~\ref{fig:halo_shapes_M15} are faded only to avoid distracting from the full physics results, not because we do not trust these results.

We show only the results for the most massive haloes, as these are the haloes in which the effects of SIDM (at least for velocity-independent cross-sections) are most pronounced. These also correspond to the haloes that can be best-studied observationally, using techniques such as weak gravitational lensing to try to infer DM halo shapes. We also looked at the shapes of $M_{200} \approx 10^{14} \msun$ haloes, finding results that were less pronounced, though qualitatively similar to the results for the high-mass haloes. As expected, and as seen for density profiles in Fig.~\ref{fig:density_profiles}, the scale on which SIDM effects are apparent decreases with decreasing halo mass, and at fixed radius the differences between CDM and SIDM are larger for more massive systems (see Fig.~\ref{fig:M200_Mdm30}). This was true also for halo shapes, which given the relatively low resolution of our simulations, and the requirement of $\sim 800$ particles to accurately measure shapes, meant that we did not resolve the scales of interest for the shapes of lower-mass haloes. At the inner-most trusted scale ($\sim 100 \kpc$ as for the $10^{15} \msun$ haloes) the median $c/a$ for our $M_{200} \approx 10^{14} \msun$ haloes ranged from 0.56 for CDM+baryons, to 0.67 for SIDM1+baryons.

\subsection{Discussion}
\label{sect:shape_discuss}

While DM self-interactions make the DM halo rounder, the stellar and gas distributions do not appear to change. This suggests that attempts to measure halo shapes using the distribution of cluster galaxies \citep{2018MNRAS.475.2421S} or the X-ray shapes of clusters \citep{2007A&A...467..485H} may struggle, at least in the context of constraints on SIDM. That changes in the DM halo shapes do not appear to be reflected in changes to the gas shapes is surprising given that an isothermal gas in hydrostatic equilibrium would have iso-density surfaces that follow the iso-potential surfaces. A detailed study of the gas properties is beyond the scope of this work, but we note here that in the inner regions the gas is likely not in hydrostatic equilibrium, with additional support from random motions or rotation \citep{2011ApJ...734...93L}, and that iso-potential surfaces are rounder than iso-density surfaces, especially in the outskirts of haloes \citep{2004IAUS..220..455J}. This means that even for CDM the gas is typically quite round, so changes to the shape of the inner regions of the DM halo may be washed out when looking at their influence on the gas shape.

Recently, \citet{2018ApJ...860L...4S} combined strong and weak lensing, X-ray photometry and spectroscopy, and the SZ effect to measure the 3D shapes of galaxy clusters. Of their 16 clusters, 11 had $c/a < 1/3$, which would clearly be at odds with any of our simulations, with or without baryons, and with any of our DM models (see the bottom-left panel of Fig.~\ref{fig:halo_shapes_M15}). They take the fact that their distribution of axis-ratios has an excess at low values over $\Lambda$CDM predictions (and a large excess over predictions including baryons) as hinting towards baryonic physics being less effective at making haloes rounder than is the case in current hydrodynamical simulations. These results would clearly be unfavourable to the SIDM hypothesis. The \citet{2018ApJ...860L...4S} model assumes the total matter distribution to be ellipsoidal, with constant axis ratios and orientation as a function of radius. This is not true of our simulated systems (see the top panels of Fig.~\ref{fig:halo_shapes_M15}), and the best-fit model with a constant axis-ratio would depend on the relative contribution of different cluster radii to the observed signal. Testing to what extent the differences in Fig.~\ref{fig:halo_shapes_M15} would show up in the analysis of \citet{2018ApJ...860L...4S} is therefore a non-trivial task, best done by generating mock X-ray, SZ and gravitational lensing observations of our simulated clusters and running them through the same pipeline as used for the real observations. We do not do this here, but remark that this would be useful both as a test of the methods used in \citet{2018ApJ...860L...4S}, and then also as a method for constraining the SIDM cross-section.


\section{Strong-lensing properties}
\label{sect:lensing}

Neither the radial density profile nor shape of a DM halo are directly observable, and inferring them from observations is fraught with difficulty (see sections \ref{sect:dens_discuss} and \ref{sect:shape_discuss}). A solution to this problem is to compare simulated haloes with real ones in terms of observed quantities, by generating mock observations from the simulations. In this section we show an example of this by generating mock strong gravitational lensing maps of our simulated clusters and comparing the properties of their critical curves with those of observed clusters. While not strictly an observable, the location of critical curves can be accurately inferred from the locations of multiply imaged background galaxies, as demonstrated for example in \citet{2017MNRAS.472.3177M}.\footnote{The majority of lens modelling techniques return an estimate of the full 2D mass distribution of the galaxy cluster, and in principle any aspect of these mass distributions could be compared with our simulations. For example, we could have compared the surface density well inside of the Einstein radius, where the differences between our different simulations would be largest, with the lens models from CLASH. However, this quantity depends sensitively on choices that went into the lens modelling, such as whether to use cored or cuspy DM haloes when performing a fit with parametric mass distributions. A recent comparison of lens modelling techniques using simulated lensing data showed that while the precise structure of the inferred critical curves varied between different lens modelling techniques, their size and overall shape are very similar across different methods, and agree with the true critical curves \citep[see Fig.~22 of][]{2017MNRAS.472.3177M}.}

\subsection{Method}

Strong gravitational lensing is the result of the gravity from a foreground mass distribution bending the path of light emitted by a background source. This can lead to background galaxies being stretched out into giant arcs, and/or being multiply imaged \citep[for a review of lensing by galaxy clusters see][]{2011A&ARv..19...47K}. Below we describe how we calculate the strong lensing properties of our simulated clusters, which involves first projecting the matter distribution onto a 2D surface density map, and then calculating the deflection angles that result from this 2D mass distribution. Finally, we use these deflection angles to calculate maps of the magnification, and focus our attention on the critical curves -- the locations of infinite magnification. The numerical parameters that we use when making our lensing maps are stated in this section without justification, but they were chosen to keep the computational cost low, while having converged results. This is discussed further in Appendix~\ref{App:lensing_tests}.

\subsubsection{2D mass maps from a particle distribution}

For each \emph{friends-of-friends} (FOF) group\footnote{For a description of the friends-of-friends algorithm, see e.g. \citet{2011ApJS..195....4M}.} from the $z=0.375$ snapshot, we start by taking all particles within $5 \, r_{200}$ of the cluster centre, defined as the location of the most bound particle. We project this material along a line of sight (the simulation $z$-axis), and generate a 2D density map, which is a square with side length $4 \mpc$ with $1024 \times 1024$ pixels, centred on the most bound particle in the cluster. We used a modified version of the triangular shaped cloud \citep[TSC,][]{1981csup.book.....H} scheme, to turn the particle distribution into a 2D density field. In standard TSC, the mass of a particle at a location $\vect{r}$ is split amongst nearby grid cells, with the cell at location $\vect{r}+\vect{x}$ receiving a fraction of the particle's mass, $W(\vect{x}) = \prod_i W(x_{i})$, with 
\begin{equation}
 W_\mathrm{TSC}(x_{i}) = 
  \begin{cases}
    0.75 - x_i^2, &  |x_i| \leq 0.5 \\
    (1.5-|x_i|)^2 / 2, & 0.5 < |x_i| \leq 1.5 \\
    0, & \text{otherwise}
  \end{cases}
\end{equation}
where $x_i$ is the $i$-th component of $\vect{x}$.

In the adaptive TSC (ATSC) scheme that we use,\footnote{Implemented in the Python package \textsc{pmesh} \citep{yu_feng_2017_1051254}.} an SPH-like smoothing length was calculated for each particle, based on the 3D distance to its 8th nearest neighbour, $r_{8}$, which is done separately for DM, gas and star particles. This distance is related to a smoothing length in pixel coordinates, $h = r_{8} / \Delta x$, where $\Delta x$ is the side-length of a pixel. The ATSC mass assignment kernel is then defined by
\begin{equation}
W_\mathrm{ATSC}(x_{i},h) = \frac{1}{h}  W_\mathrm{TSC}(x_{i}/h).
\end{equation}
This breaks down for $h<1$, and so we set a minimum of $h=1$. Also, large $h$ leads to the particle mass being split between many grid cells, which is computationally expensive; we therefore set a maximum of $h=10$.

We use the ATSC scheme to make separate maps of the DM, stars and gas, and sum these together with a black hole map to get a total projected-density map, $\Sigma(x,y)$. For the black hole map we use standard TSC rather than ATSC, i.e. we set $h=1$ for all black holes.

\begin{figure*}
        \centering
        \includegraphics[width=\textwidth]{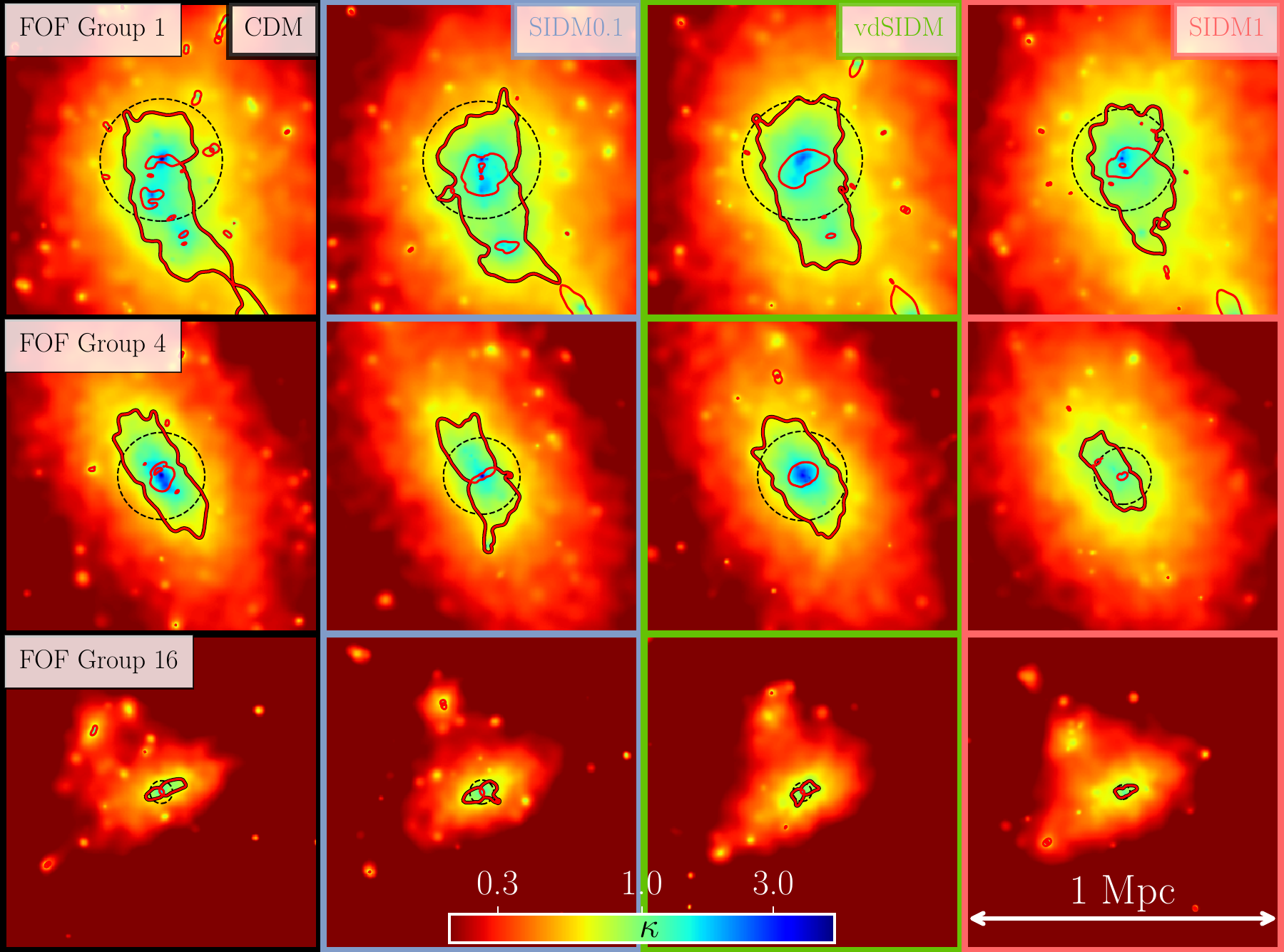}
	\caption{Convergence maps of the 1st, 4th and 16th most massive FOF groups in the BAHAMAS simulations, simulated with different SIDM cross-sections. The critical curves (with a lens redshift of $z_\mathrm{l} = 0.375$ and a source redshift $z_\mathrm{s} = 2$) are plotted in red, with the largest tangential critical curve having a black outline. The dashed circles contain the same area as this largest tangential critical curve, and therefore represent the effective Einstein radius, $\effein$. The scale of each panel is the same, covering a field of view of $1 \mpc$ in the lens plane, corresponding to approximately 3.2 arcmin.}
	\label{fig:strong_lensing_maps}
\end{figure*}

\subsubsection{Ray-tracing through a mass distribution}

The projected-density map, is then scaled by the critical surface-density for lensing, to produce a convergence  map
\begin{equation}
\kappa(x,y) = \frac{\Sigma(x,y)}{\Sigma_\mathrm{crit}},
\label{kappa_crit}
\end{equation}
where $\Sigma_\mathrm{crit}$ depends on the geometry of the source, observer and lens through
\begin{equation}
\Sigma_\mathrm{crit} = \frac{c^2}{4 \pi G} \frac{D_\mathrm{s}}{D_\mathrm{l}D_\mathrm{ls}}.
\label{sigma_crit}
\end{equation}
Here, $D_\mathrm{s}$, $D_\mathrm{l}$, and $D_\mathrm{ls}$ are the angular diameter distances between the observer and the source, observer and lens, and lens and source respectively. These in turn depend on the cosmological parameters and redshifts of the lens (i.e. the simulated galaxy cluster) and source. We used the same WMAP9 cosmology \citep{2013ApJS..208...19H} as used to run the simulations, with an assumed source redshift of $z_\mathrm{s} = 2$, and lens redshift $z_\mathrm{l} = 0.375$. 

The convergence field is related to the effective lensing potential, $\Psi$, through
\begin{equation}
\kappa = \frac{1}{2} \nabla^2 \Psi \equiv \left(\frac{\partial^2 \Psi}{\partial x^2} + \frac{\partial^2 \Psi}{\partial y^2} \right).
\label{kappa_psi}
\end{equation}
The deflection angle field can also be related to the effective potential,
\begin{equation}
\vect{\alpha} = \vect{\nabla} \Psi \equiv \left( \pdv{\Psi}{x} , \pdv{\Psi}{y} \right).
\label{alpha_psi}
\end{equation}
Equations \eqref{kappa_psi} and \eqref{alpha_psi} lead to a simple relationship between the Fourier transforms of $\kappa$ and $\vect{\alpha}$ ($\hat{\kappa}$ and $\hat{\vect{\alpha}}$ respectively), namely
\begin{equation}
\hat{\vect{\alpha}} = \frac{2 i \hat{\kappa}}{|\vect{k}|^2} \vect{k},
\label{alpha_kappa}
\end{equation}
where $\vect{k} = (k_x, k_y)$ is the wave vector conjugate to $\vect{x} = (x, y)$. This allows us to efficiently generate $\vect{\alpha}(x,y)$ on the same regular grid as $\kappa(x,y)$, by performing a discrete Fourier transform on $\kappa$ to get $\hat{\kappa}$, using equation \eqref{alpha_kappa} to get $\hat{\vect{\alpha}}$, and then taking the inverse discrete Fourier transform of $\hat{\vect{\alpha}}$ to get $\vect{\alpha}$. The discrete Fourier transform implicitly assumes the function to be periodic, which is not the case for the convergence field of an isolated cluster. To reduce the error caused by this, we surrounded the cluster by a zero-padded field out to $4096 \times 4096$ (i.e. increasing by a factor of four along both axes).

\subsubsection{Calculating observable properties}
\label{sect:lensing_observables}

The magnification and distortion of a background source can be computed from the Jacobian matrix, $\mathbfss{A}$, of the mapping from the unlensed to lensed coordinate systems. It can be written in terms of gradients of the deflection angle field 
\begin{equation}
\mathbfss{A} = \delta_{ij} - \pdv{\alpha_i(\vect{x})}{x_j}.
\label{jacobian}
\end{equation}
We interpolate $\vect{\alpha}(\vect{x})$ onto a finer $2048 \times 2048$ grid,\footnote{We make use of the \textsc{scipy} \citep{scipy} function RectBivariateSpline, to do bicubic interpolation.} in the central $2 \mpc$ of the field. We then use a finite difference method to find the derivates of $\vect{\alpha}(\vect{x})$ on this finer grid, from which we construct $\mathbfss{A}(\vect{x})$.

The magnification is given by the inverse of the determinant of the Jacobian matrix,
\begin{equation}
\mu = \frac{1}{\det \mathbfss{A}}.
\label{eq:mag}
\end{equation}
The \emph{critical lines} are regions in the lensing plane where $\det \mathbfss{A} = 0$ and the magnification is formally infinite. Critical lines come in two varieties, known as \emph{radial} and \emph{tangential}. Writing the Jacobian matrix as
\begin{equation}
\mathbfss{A} = 
\begin{pmatrix} 1-\kappa-\gamma_1 & -\gamma_2\\ -\gamma_2 & 1-\kappa+\gamma_1 \end{pmatrix},
\label{eq:lensing_Jacobian}
\end{equation}
with $\vect{\gamma}$ a pseudo-vector known as the shear
\begin{equation}
\vect{\gamma} = (\gamma_1 , \gamma_2) \equiv \left( \frac{1}{2} \left( \pdv[2]{\Psi}{x} -  \pdv[2]{\Psi}{y} \right) , \pdv{\Psi}{x}{y} \right),
\label{eq:shear}
\end{equation}
we see that the determinant of $\mathbfss{A}$ can be written as
\begin{equation}
\det \mathbfss{A} = \left( 1 - \kappa - \gamma \right) \left( 1 - \kappa + \gamma \right), 
\label{eq:det_A}
\end{equation}
where $\gamma = |\vect{\gamma}|$. \emph{Radial critical lines} appear where $1 - \kappa + \gamma = 0$, with images close to this line stretched in the direction perpendicular to the line. \emph{Tangential critical lines} occur where $1 - \kappa - \gamma = 0$, and lead to images stretched tangentially to the line. For axisymmetric lenses, the latter of these correspond to the \emph{Einstein ring}, and it is these tangential critical curves that are the main focus of the rest of this section.

There are of course many ways in which the lensing properties of our simulated clusters could be compared with observed systems. The location of tangential critical curves is one that can be well-constrained observationally \citep[e.g.][]{2010A&A...514A..93M, 2013SSRv..177...75H} owing to the bulk of multiple image systems lying close to these curves. An axisymmetric lens will have a circular critical curve, and the Einstein radius, $\theta_\mathrm{E}$, is defined as the angular radius of this circle. Extending the definition of the Einstein radius to cases where the critical curves are no longer circles can be done in numerous ways, with a good overview of previously used methods in \citet{2013SSRv..177...31M}. We choose to use the \emph{effective Einstein radius}, $\theta_\mathrm{E,eff}$, because it correlates tightly with the probability of producing giant lensing arcs, and is less sensitive to cluster mergers than other definitions \citep{2012A&A...547A..66R}. This is defined as 
\begin{equation}
\theta_\mathrm{E,eff} = \sqrt{\frac{A}{\pi}}, 
\label{eq:thetaEeff}
\end{equation}
where $A$ is the area enclosed by the tangential critical curve. Clearly for the case of a circular critical curve this definition agrees with the definition of $\theta_\mathrm{E}$.

From each of our lensing maps, we extract the tangential critical curves ($1 - \kappa - \gamma = 0$ contours), using a \emph{marching squares} method.\footnote{Implemented as \code{find\_contours} in scikit-image \citep{scikit-image}.} Each lensing map can have multiple components with their own tangential critical curves, but we take only the longest closed tangential critical curve within each map. While this critical curve generally encloses the halo centre,\footnote{For CDM (SIDM1), this is true for over 85 (60) percent of haloes with $M_{200} > 10^{14} \msun$.} we do not enforce that it does so. The area enclosed by this curve (defined by a set of points on the curve) is calculated using Green's theorem. Specifically, the area is calculated from an integral along the entirety of the critical curve:  
\begin{equation}
A = \iint \mathrm{d}x \, \mathrm{d}y =  \oint x \mathrm{d}y.
\label{eq:GreensTheorem}
\end{equation}

Accurately mapping out the critical curves of our clusters requires that the mass distribution in the inner region of the halo is well-sampled. Owing to the finite resolution of both our simulations and our lensing procedure, we therefore expect there to be some minimum halo mass, below which we cease to trust our results. By running our lensing analysis on mass distributions that sub-sampled particles from the simulations, we found that whether or not $\theta_\mathrm{E,eff}$ had converged depended more on $\theta_\mathrm{E,eff}$ itself than the mass of the halo. We discuss this further in Appendix~\ref{App:lensing_tests}, but note here that we expect our $\theta_\mathrm{E,eff}$ values to be converged when $\theta_\mathrm{E,eff} \gtrsim 2 \, \mathrm{arcsec}$.

\subsection{Results}
\label{sect:lensing_results}

In Fig.~\ref{fig:strong_lensing_maps} we show maps of the lensing convergence of three fairly massive galaxy clusters for each of our DM models, with the critical curves overlaid. The haloes we show are the first, fourth and sixteenth most massive FOF groups from the simulation (at $z=0.375$), with virial masses ($M_{200}$) of $\sim 1.7, 1.5$ and $0.6 \times 10^{15} \msun$ respectively.\footnote{Note that while the $M_{200}$ values of haloes change depending on the cross-section, there does not seem to be a systematic shift, and these changes are typically only at the percent level.}

\begin{figure}
        \centering
        \includegraphics[width=\columnwidth]{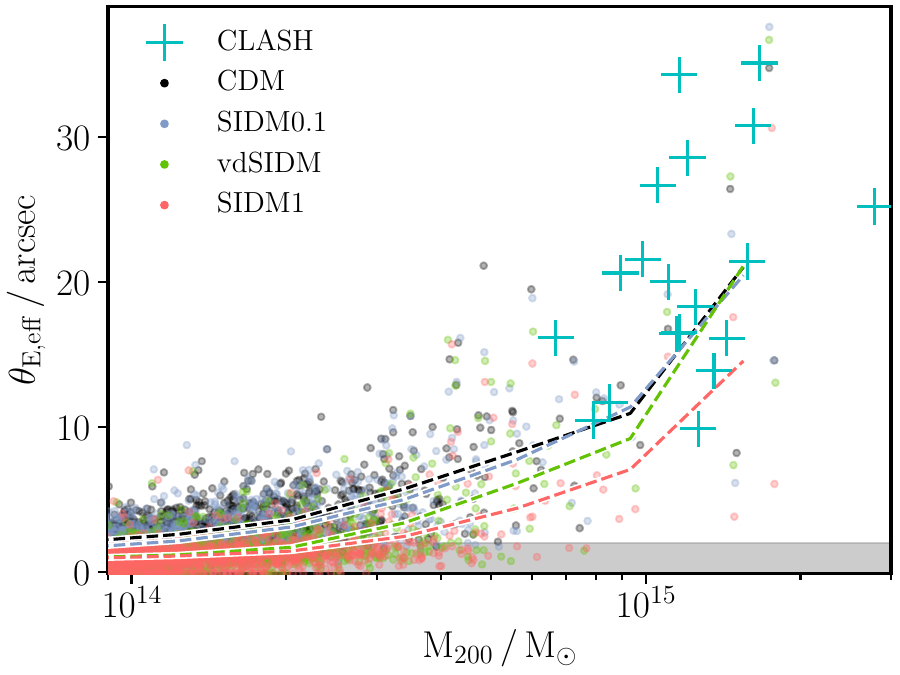}
	\caption{The effective Einstein radii of all simulated clusters with $M_{200} > \num{9e13} \msun$, for our four different DM models. The clusters were taken from the $z=0.375$ snapshot, which was the lens redshift used, while the source redshift was $z_\mathrm{s} = 2$. The dashed lines show the median $\effein$ as a function of halo mass, with the points showing individual haloes. The grey shaded region indicates where we expect $\effein$ to be under-estimated due to the resolution of our simulations (discussed in Appendix~\ref{App:lensing_tests}). The cyan crosses show CLASH clusters scaled to be at $z=0.375$ (see Section~\ref{sect:clash}).}
	\label{fig:M200-equiveinstein}
\end{figure}

In Fig.~\ref{fig:M200-equiveinstein} we plot $\theta_\mathrm{E,eff}$ against halo mass for haloes at $z_\mathrm{l}=0.375$ with a source plane at $z_\mathrm{s} = 2$. While there is substantial scatter in $\theta_\mathrm{E,eff}$ at fixed $M_{200}$ within each DM model, there are clear shifts in the $\theta_\mathrm{E,eff}$ distributions as the DM model is changed. A change in the distribution of Einstein radii with SIDM+baryons compared with CDM+baryons was also recently seen in simulated haloes with $M_{200} \sim 10^{13} \msun$ by \citet{2019MNRAS.484.4563D}, though at that mass scale the dependence of Einstein radius on DM model is complicated because the strong lensing regions are more baryon dominated than in clusters.

\begin{figure}
        \centering
        \includegraphics[width=\columnwidth]{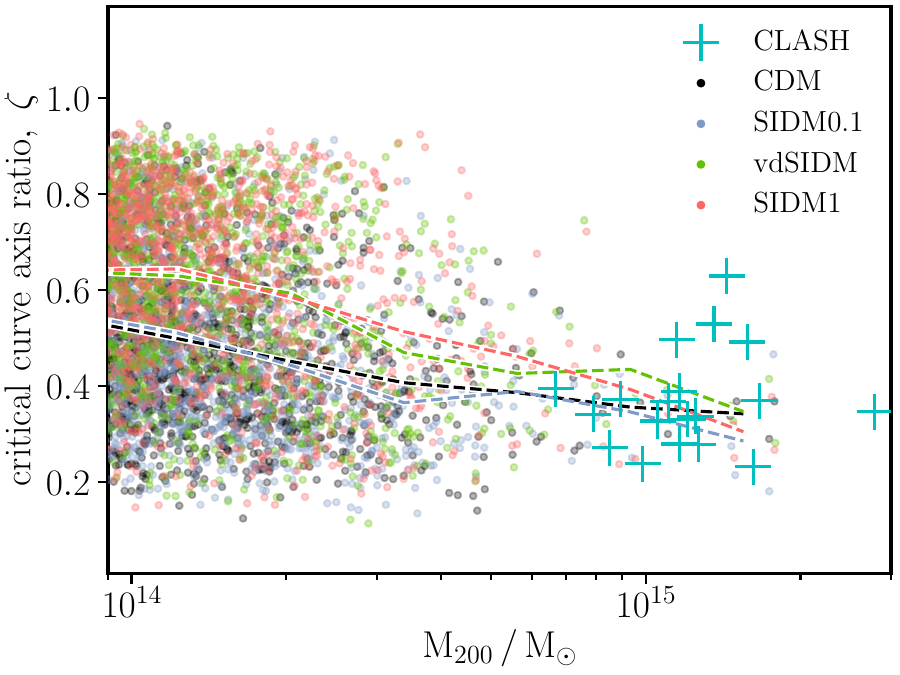}
	\caption{The minor-to-major axis ratios of our simulated clusters' critical curves, modelling them as ellipses (described in Section~\ref{sect:lensing_results}). The dashed lines are median axis ratios as a function of $M_{200}$ for the different DM models. The cyan crosses show CLASH clusters scaled to be at $z=0.375$ (see Section~\ref{sect:clash}).}
	\label{fig:M200-crit-curve-axis-ratio}
\end{figure}

Aside from changes to the radial density profile, SIDM tends to make the centre of haloes rounder (as discussed in Section~\ref{sect:halo_shapes}). It might, therefore, be expected that this roundness is reflected in the critical curves, and so we calculate their \emph{axis ratios}. Defining the furthest distance between two points on the critical curve to be $l_\mathrm{max}$, we then define the axis ratio as
\begin{equation}
\zeta = \frac{4 \, A}{\pi \, l_\mathrm{max}^2} \, ,
\label{eq:crit_curve_axis_ratio}
\end{equation}
where A is still the area enclosed by the critical curve. For an elliptical critical curve, this definition of axis ratio is the ratio of the semi-minor and semi-major axes. We show this axis ratio as a function of halo mass for our different DM models in Fig.~\ref{fig:M200-crit-curve-axis-ratio}. Haloes generally have rounder critical curves with larger SIDM cross-sections. More massive haloes have more elongated critical curves. These two effects unfortunately conspire such that at fixed $\effein$ there is no clear trend between the DM model and $\zeta$ (not shown). In other words, the tangential critical curve of an SIDM halo of a given mass looks similar (in terms of its size and roundness) to that of a CDM halo that is less massive. While this precludes a strong-lensing-only test of the DM model, if combined with an appropriate $M_{200}$ measurement then $\effein$ and $\zeta$ can both be used to constrain the DM model.

\subsubsection{CLASH clusters}
\label{sect:clash}

For comparison with our simulated clusters, we used clusters from the CLASH survey \citep{2012ApJS..199...25P}. Five of the twenty-five CLASH clusters were selected because of their extreme gravitational lensing properties, making them a highly biased sample that we exclude from our analysis. Of the other twenty, which were selected based on X-ray luminosity, one had no wide-field weak lensing data suitable for estimating $M_{200}$. We therefore use the same sample of nineteen X-ray selected clusters as used in \citet{2015ApJ...806....4M}, taking our $M_{200}$ estimates from this paper also. We assume the X-ray selection criteria used are a suitable proxy for mass-selection, such that the CLASH clusters have unbiased $\effein$ at a given $M_{200}$. This may not be strictly true,\footnote{One could certainly imagine that at fixed halo mass: more centrally concentrated clusters, with larger $\effein$, are also brighter in the X-ray.} and we would ideally apply the CLASH selection function to our simulated clusters. However, at the high-mass end probed by CLASH we have very few simulated clusters, and cannot meaningfully replicate the CLASH selection.

For the strong-lensing properties of the CLASH clusters we used the \emph{PIEMD + eNFW} mass models constructed by \cite{2015ApJ...801...44Z}, using the method from \citet{2013ApJ...762L..30Z}. These were obtained through the \emph{Hubble Space Telescope} Archive as a high-end science product of the CLASH program. The CLASH sample spans a range of redshifts from 0.19 to 0.89. In order to compare directly with our simulated clusters, we re-scale the lensing maps to common lens and source redshifts of $z_\mathrm{l} = 0.375$ and $z_\mathrm{s} = 2$. This involves taking the CLASH convergence and shear maps (expressed on a grid of angular coordinates), and multiplying their normalisation by the ratio of the critical surface density used to generate them to the critical surface density for our chosen lensing geometry, and then also re-scaling the angular coordinates by the ratio of the angular diameter distance to the CLASH cluster to the angular diameter distance to $z_\mathrm{l}=0.375$. Once we have re-scaled the convergence and shear maps, we follow the procedure described in Section~\ref{sect:lensing_observables} to calculate the tangential critical curves. We apply a similar re-scaling to the $M_{200}$ values from \citet{2015ApJ...806....4M}. We take their best-fit physical parameters that describe the halo ($\rho_\mathrm{s}$ and $r_\mathrm{s}$ of NFW profiles), and calculate the corresponding value of $M_{200}$ at $z=0.375$.

\subsection{Discussion}

Although large by the standards of hydrodynamical simulations, the box size of \bahamas does not produce many clusters of comparable mass to the CLASH sample. As such, it is difficult to make firm statements about the SIDM cross-section from this comparison. Nevertheless, the fact that only one CLASH cluster lies below the median line for SIDM1 in Fig.~\ref{fig:M200-equiveinstein} suggests that $\sigma/m < 1 \cmsg$ at velocities of order $1000 \kms$. A better comparison would either require a much larger volume simulation, or dedicated zoom simulations of a reasonable number of CLASH-like clusters. Alternatively, an observed cluster sample at lower masses would have a substantial number of simulated counterparts. Unfortunately, these lower-mass systems have smaller $\effein$ and correspondingly smaller cross-sections for producing lensed images. This means that without exceptionally deep data, most systems with lower masses will not produce enough lensed images for their strong-lensing properties to be well-constrained, with those that do suffering a bias towards being the systems with the largest $\effein$.

Another limitation of comparing our simulated clusters with the observed CLASH clusters to constrain SIDM is our uncertainty surrounding a `correct' implementation of baryonic physics. We show in Appendix~\ref{App:subgrid} that CDM simulations with AGN feedback model parameters spanning the observationally allowed range are more similar in their lensing properties than CDM and SIDM1 with common AGN feedback parameters. If the effects of AGN in an SIDM1 universe are the same as their effects in a CDM universe (i.e. they add to or multiply the Einstein radii by the same amount), then this result will allow us to distinguish between CDM and SIDM1, even with current uncertainty surrounding subgrid baryonic physics. However, it is possible that less efficient AGN feedback, which leads to more massive central galaxies, will have a more pronounced effect in an SIDM rather than CDM universe (as seen in the differences between DM-only and full physics density profiles in the middle panels of Fig.~\ref{fig:density_profiles}). Here we stress that our forecasts are contingent upon the fiducial BAHAMAS model being an accurate description of galaxy formation physics on the scales of interest.

The fact that the simulated clusters produce critical curves is an interesting result by itself. Using SIDM-only simulations of a galaxy cluster, \citet{2001MNRAS.325..435M} found that even moderate cross-sections ($\sigma/m \sim 0.1 \cmsg$) led to galaxy clusters incapable of producing critical curves. Ray-tracing our SIDM-only simulations we find results that agree with \citet{2001MNRAS.325..435M}, with the bulk of vdSIDM-only and SIDM1-only systems producing no critical curves, and SIDM0.1-only haloes having substantially smaller $\theta_\mathrm{E,eff}$ than CDM-only, or often not having any critical curves either. When including baryons, both the baryonic mass distribution itself and the effect it has on the DM distribution substantially revise this to the point where CDM and SIDM0.1 have very similar strong-lensing properties, and even SIDM1 produces substantial critical curves, albeit ones that appear inconsistent with the CLASH sample.

\section{Conclusions}
\label{sect:conclusions}

SIDM has become an attractive alternative to CDM, due to its ability to decrease the DM density at the centre of dwarf galaxies, potentially bringing simulated dwarf galaxies into better agreement with observed ones. SIDM cross-sections that produce significant effects on dwarf galaxy scales would strongly affect galaxy clusters, unless the cross-section decreases with increasing DM--DM velocity. We have therefore simulated the first large cosmological volumes with both SIDM and baryonic physics, to enable robust constraints on SIDM from cluster scales. Our simulations used the galaxy formation code used for the \bahamas project \citep{2017MNRAS.465.2936M}, which was specifically calibrated to reproduce observables relevant to galaxy groups and clutsers, including their gas fraction and the high-mass end of the stellar mass function. 

We have shown that density profiles are substantially affected by the inclusion of baryons, with SIDM haloes affected by baryons out to larger radii than their CDM counterparts. However, the relative increase in DM density when including baryons is not especially large with SIDM compared with CDM (Fig.~\ref{fig:density_profiles}). We find that the \emph{diverse density profiles} found for two high-resolution SIDM \ceagle cluster simulations presented by \citet{2018MNRAS.476L..20R} are most likely a result of being (un)lucky with a small sample size, or driven by the large stellar masses of their central galaxies. That being said, we cannot rule out that the lower-resolution simulations shown here do not resolve the appropriate times and scales necessary to produce this diversity.

One of our simulated SIDM cross-sections was velocity-dependent. For this cross-section, the density profiles of haloes of a given mass can be approximately reproduced using a velocity-independent and isotropic cross-section. This works provided that the velocity-independent cross-section is equal to the momentum transfer cross-section of the velocity-dependent one at a velocity scale typical of the halo in question ($\sim \sqrt{G\, M_{200} / r_{200}}$). 

The DM haloes of our simulated clusters were made rounder by SIDM out to scales comparable with the virial radius (Fig.~\ref{fig:halo_shapes_M15}). The effect of including baryons was roughly independent of the DM model, and the differences between CDM and our largest SIDM cross-section were substantially larger than the differences between DM-only and DM+baryons. Interestingly, while the DM haloes were rounder with SIDM, this was not reflected in either the gas or stellar distributions, suggesting that lensing may be the best way to probe the effects of SIDM on the shapes of haloes.

Density profiles and halo shapes are not directly observable, and inferring them from observations can be difficult. Especially for the case of halo shapes, there are many different definitions used when analysing simulations, and none of them map precisely on to what one would measure observationally. The density profiles and halo shapes we show are therefore merely illustrations of an effect, providing an indication of the magnitude of expected changes when changing DM model. A comparison with observational data will require the same method to also be run on simulations, for which full hydrodynamical simulations -- as presented here -- are ideal, because mock data sets can be generated from the different particle species simulated.

We provided one example of generating observable (or at least close to observable) quantities from our simulated clusters, producing strong gravitational lensing maps of them (for examples, see Fig.~\ref{fig:strong_lensing_maps}) and analysing the properties of their critical curves. While we did not have a large number of simulated clusters of comparable mass with well-studied observed clusters, our results suggest that $\sigma/m = 1 \cmsg$ is in slight tension with observed cluster strong lensing. This was because a cross-section of this size leads to clusters with smaller Einstein radii at a given halo mass than is observed (Fig.~\ref{fig:M200-equiveinstein}). We note that this constraint is substantially weaker than the \citet{2001MNRAS.325..435M} constraint of $\sigma/m \lesssim 0.1 \cmsg$, because this earlier constraint relied on SIDM-only simulations for which moderate cross-sections lead to an absence of critical curves. This highlights the importance of including baryons in simulations with SIDM, both because of the direct effect of the baryonic distribution on observable quantities, as well as their indirect effect through influencing the structure of DM haloes.

\section*{Acknowledgments}

AR is supported by the European Research Council (ERC-StG-716532-PUNCA) and the STFC (ST/N001494/1). DH acknowledges support by the Merac foundation. RM acknowledges the support of a Royal Society University Research Fellowship. VRE acknowledges support from STFC grant ST/P000541/1. BL is supported by an European Research Council Starting Grant (ERC-StG-716532-PUNCA) and STFC (ST/L00075X/1, ST/P000541/1).

This work used the DiRAC Data Centric system at Durham University, operated by the Institute for Computational Cosmology on behalf of the STFC DiRAC HPC Facility (www.dirac.ac.uk). This equipment was funded by BIS National E-infrastructure capital grant ST/K00042X/1, STFC capital grants ST/H008519/1 and ST/K00087X/1, STFC DiRAC Operations grant  ST/K003267/1 and Durham University. DiRAC is part of the National E-Infrastructure. This project has received funding from the European Research Council (ERC) under the European Union's Horizon 2020 research and innovation programme (grant agreement No 769130). This research is supported by the Swiss National Science Foundation (SNSF).

\bibliographystyle{mnras}
\bibliography{bibliography}

\appendix

\section{The impacts of different subgrid models}
\label{App:subgrid}

As discussed in Section~\ref{sect:BAHAMAS_physics}, our simulations use subgrid models to implement the baryonic physics associated with galaxy formation. These models contain free parameters, which in the case of \bahamas were adjusted so that the simulations adequately reproduced the observed present-day galaxy stellar mass function and the hot gas mass within groups and clusters of galaxies. A full investigation of the impact that different subgrid model choices would have had on this work is beyond its scope. However, in this Appendix we give an indication of the sensitivity of our results to the subgrid model used  by showing the results of CDM simulations with a couple of different subgrid model parameter choices.

\begin{figure}
        \centering
        \includegraphics[width=\columnwidth]{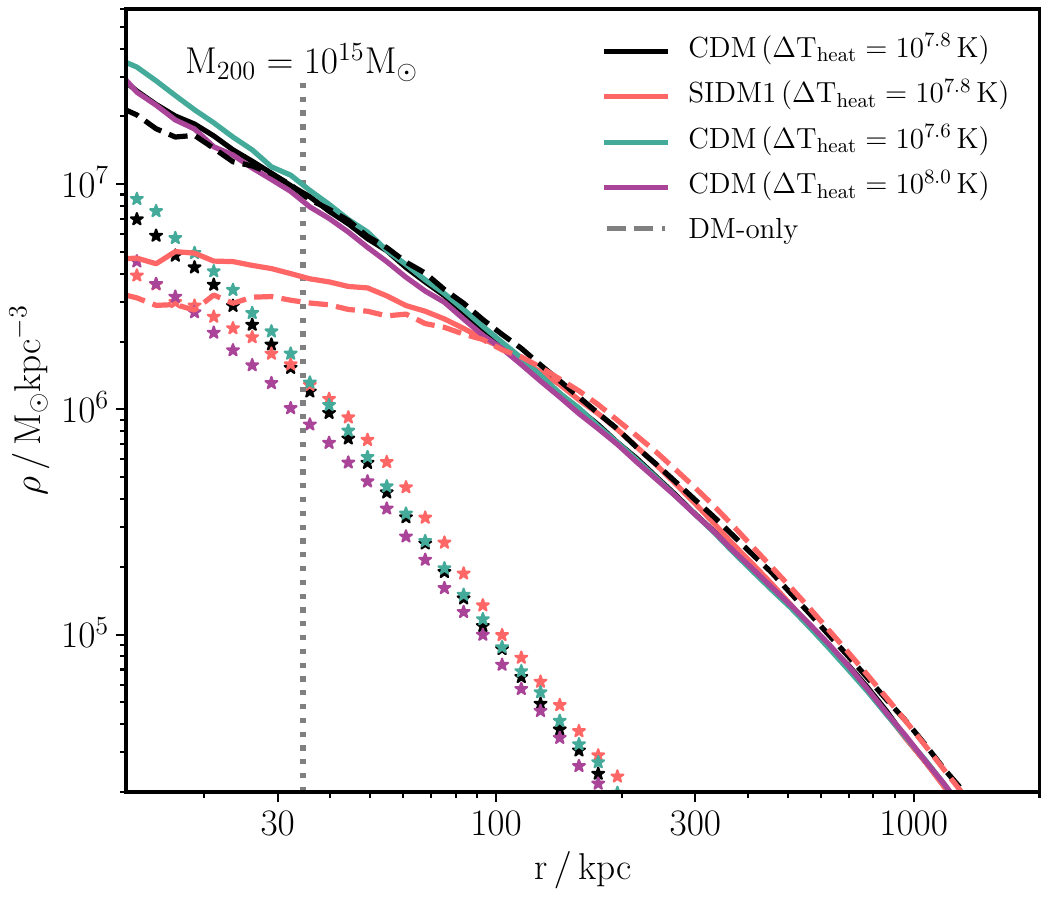}
	\caption{The same as the density panel from the right column of Fig.~\ref{fig:density_profiles} but plotting only CDM and our largest SIDM cross-section, and additionally showing the density profiles of two additional CDM models with different AGN heating temperatures. While the changes to the stellar density induced by changing DM model or changing subgrid physics parameters are comparable, the DM density profile is more sensitive to the DM model than to changing $\Delta T_\mathrm{heat}$.}
	\label{fig:AGN_density_profiles}
\end{figure}

\begin{figure}
        \centering
        \includegraphics[width=\columnwidth]{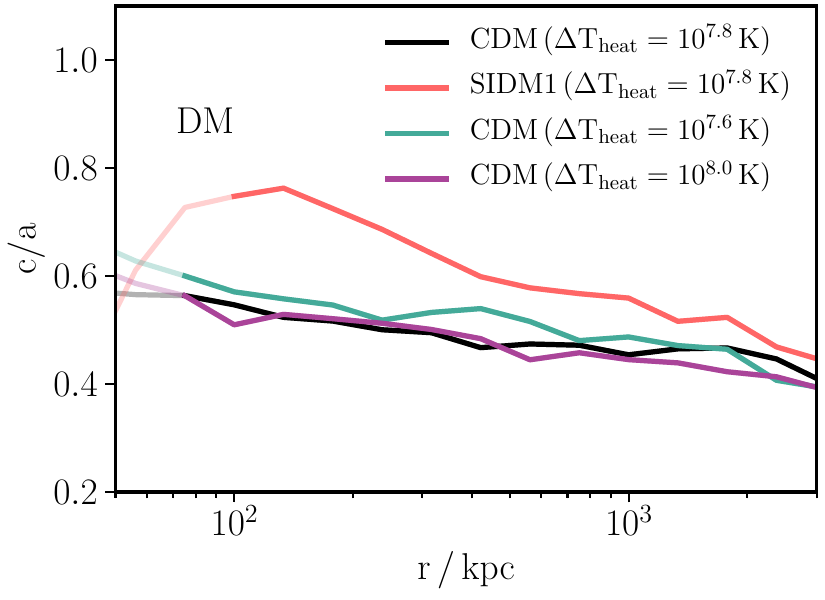}
	\caption{The same as the top-left panel of Fig.~\ref{fig:halo_shapes_M15} but including the halo shapes from two additional CDM models with different AGN heating temperatures. The increase in median $c/a$ with SIDM1 is substantially larger than the changes when varying $\Delta T_\mathrm{heat}$.}
	\label{fig:AGN_halo_shapes}
\end{figure}

\begin{figure}
        \centering
        \includegraphics[width=\columnwidth]{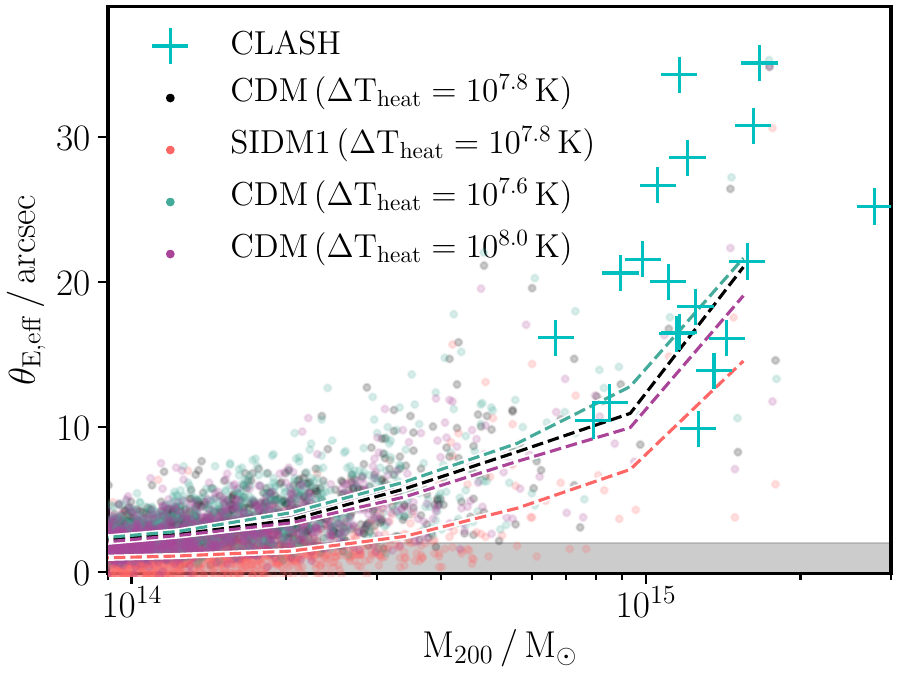}
	\caption{The same as Fig.~\ref{fig:M200-equiveinstein} but including the effective Einstein radii from simulated clusters with two additional CDM models with different AGN heating temperatures. The slight changes in median $\effein$ with different $\Delta T_\mathrm{heat}$ reflect the changes to the stellar and DM densities seen in Fig.~\ref{fig:AGN_density_profiles}.}
	\label{fig:AGN_einstein_radii}
\end{figure}

In Figures \ref{fig:AGN_density_profiles}, \ref{fig:AGN_halo_shapes} and \ref{fig:AGN_einstein_radii} we compare the size of effects from changing the specifics of the subgrid galaxy formation physics with those from changing DM model. Specifically, we compare three different CDM+baryons simulations with SIDM1+baryons. These different CDM+baryons simulations vary the temperature to which gas is heated by AGN. The fiducial model that was used throughout the rest of this paper increases the temperature of AGN-heated particles by $\Delta T_\mathrm{heat} = 10^{7.8} \,$K. The two alternative models that we consider in this appendix have $\Delta T_\mathrm{heat} = 10^{7.6}$ and $10^{8.0}\,$K. A larger value of $\Delta T_\mathrm{heat}$ makes the AGN feedback stronger. \citet{McCarthy2018} showed that these two models bracket the observed cluster gas fractions.

We find that the DM density profiles of $M_{200} \approx 10^{14} \msun$ clusters are unaffected by $\Delta T_\mathrm{heat}$, while the stellar densities decrease with increasing $\Delta T_\mathrm{heat}$ (not shown). In systems with $M_{200} \approx 10^{15} \msun$ the changes in stellar density are more pronounced, and the DM profiles are also affected. The mean density profiles of these more massive systems are shown in Fig.~\ref{fig:AGN_density_profiles}, where it can be seen that changing the DM model from CDM to SIDM1 leads to a substantially larger change in the DM density profile than varying the subgrid physics (keeping the DM model fixed).

We also investigated how the halo shapes are affected by changing $\Delta T_\mathrm{heat}$. We found that the DM haloes were slightly more spherical with $\Delta T_\mathrm{heat} = 10^{7.6} \,$K than the other heating temperatures, and that the gas and stars were not noticeably affected. In Fig.~\ref{fig:AGN_halo_shapes} we plot the minor-to-major axis ratios of the DM haloes with different $\Delta T_\mathrm{heat}$, showing that the halo shape changes induced by an SIDM cross-section of $1 \cmsg$ are considerably larger than those from the changes we make to the subgrid model parameters.

Finally, in Fig.~\ref{fig:AGN_einstein_radii} we plot the effective Einstein radii of galaxy clusters with CDM and SIDM1 (with our fiducial value of $\Delta T_\mathrm{heat} = 10^{7.8} \,$K with the two alternative CDM runs. In a similar manner to the density profiles and halo shapes, the CDM models with different $\Delta T_\mathrm{heat}$ are more similar than CDM and SIDM1 with a common $\Delta T_\mathrm{heat}$. The same is not true, however, if we compared going from CDM to SIDM0.1 with the different $\Delta T_\mathrm{heat}$ values. This means that while uncertainty in the baryonic physics is probably not a major hurdle for constraining $\sigma/m \lesssim 1 \cmsg$, future experiments that hope to lower this bound will have to consider how different models for galaxy formation that span the uncertainties on relevant observables, as is the case for the models considered here, might affect their results.

\section{Convergence of halo shapes}
\label{App:halo_shape_tests}

In order to check where our halo shape measurements (discussed in Section~\ref{sect:halo_shapes}) can be trusted, we tested our shape measurement algorithm on particle distributions with known axis ratios. We start by noting that measuring a halo shape with only a small number of particles will tend to lead to more extreme axis ratios, as the particle noise will lead to some axis along which the distribution appears elongated. To take a simple example, the mean minor-to-major axis ratio derived from the reduced inertia tensor (equation~\ref{mass_tensor}) calculated from $N$ particles placed randomly on the unit sphere is 0.54, 0.85, 0.95 and 0.98 for $N=10$, 100, 1000 and $10\,000$ respectively.

\begin{figure*}
        \centering
        \includegraphics[width=\textwidth]{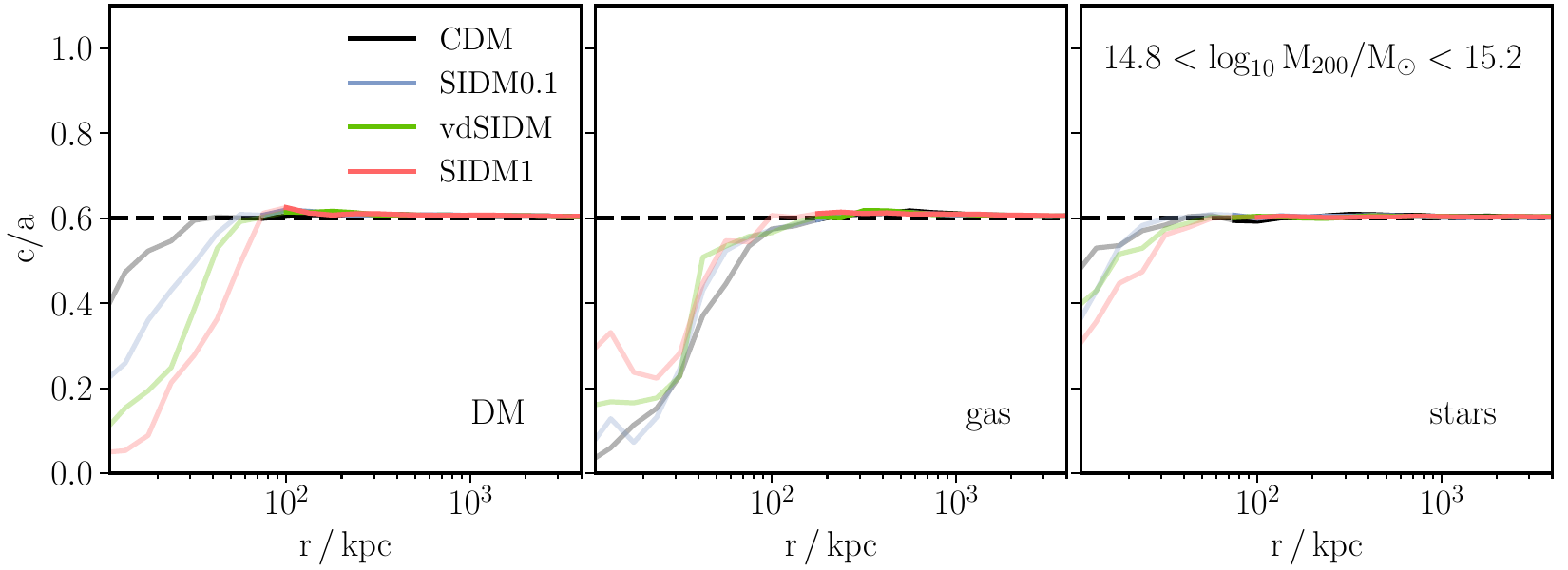}
	\caption{The median minor-to-major axis ratios of \emph{shuffle-stretched} haloes, with a known true axis ratio of 0.6. The method used to derive halo shapes and generate the plot was identical to Fig.~\ref{fig:halo_shapes_M15}, but with the particle distributions from the simulated haloes altered following the procedure described in Appendix~\ref{App:halo_shape_tests}. The lines are semi-transparent at radii where the median number of particles used for the shape measurement is less than 800. At small radii, where a small number of particles contribute to the shape measurement, the method is biased towards returning lower $c/a$ than the true value. This bias is not present when more than 800 particles are used for the shape measurement. }
	\label{fig:halo_shapes_M15_shuffle}
\end{figure*}

While we use the \emph{reduced} inertia tensor (with $1/r_{n}^2$ weighting), to which particles at different radii contribute roughly equally, our method is still sensitive to the radial distribution of particles. This is because during the iterative procedure only particles near the boundary of our ellipsoidal search region enter and leave the search region as it is deformed. In order to test our halo shape algorithm on particle distributions with known axis ratios and relevant radial distributions of particles (which differ for the different DM cross-sections) we take simulated haloes and turn them into \emph{shuffle-stretched} versions. What this means is that we take all particles within a halo and first re-distribute each particle randomly on a sphere (drawing uniformly in $\cos \theta$ and $\phi$), keeping its distance from the halo centre constant. We then stretch the haloes to have axis ratios $s \equiv  c/a=0.6$ and $q \equiv b/a=0.8$, roughly representative of the axis ratios of our real haloes (see Fig.~\ref{fig:halo_shapes_M15}). Specifically, we stretch in the $x$, $y$ and $z$ directions by factors of $(s\,q)^{-1/3}$, $q^{2/3} \, s^{-1/3}$ and $s^{2/3} \, q^{-1/3}$ respectively, keeping the ellipsoidal radius (equation \ref{elliptical_radius}) of each particle equal to its initial spherical radius. We then run an identical procedure as used for the top panel of Fig.~\ref{fig:halo_shapes_M15} to make Fig.~\ref{fig:halo_shapes_M15_shuffle}, using the \emph{shuffle-stretched} haloes.

We find that at small radii, where a low number of particles contribute to the shape measurement, $c/a$ is systematically underestimated. We found with all cross-sections, and for all three different matter species, that the relative error in the median $c/a$ is never larger than 5\% so long as at least 800 particles are used. We therefore make the lines in both Fig.~\ref{fig:halo_shapes_M15} and Fig.~\ref{fig:halo_shapes_M15_shuffle} semi-transparent at radii where the median number of particles contributing to the shape measurement is fewer than 800.

At large $r$ there is a slight bias towards measuring haloes to be more spherical than they truly are, with median values of $c/a$ at large radii closer to 0.61 than 0.6. We experimented with varying $\epsilon_\mathrm{conv}$ (equation \ref{halo_shape_convergence}), and found that this bias increased systematically with increasing $\epsilon_\mathrm{conv}$. The reason is that our iterative procedure starts from a sphere, and so we typically converge on a value of $c/a$ from above. A finite value of $\epsilon_\mathrm{conv}$ means that this procedure stops before it has fully converged. Using a smaller $\epsilon_\mathrm{conv}$ requires a larger number of iterations, increasing the required computational resources, and can also result in a number of haloes failing to converge.\footnote{This can happen if, for example, the eigenvalues and eigenvectors of the inertia tensor from iteration $i$ leads to a search volume for iteration $i+1$ that includes the same particles as used in iteration $i-1$ (without reaching convergence in $q$ and $s$), such that the iteration gets stuck in a loop switching between two sets of particles with their associated inertia tensors. While we saw this phenomenon at small radii (with few particles), none of our haloes failed to converge at the radii we trust (where lines are non-transparent in Fig.~\ref{fig:halo_shapes_M15}).} Our fiducial value of $\epsilon_\mathrm{conv} = 0.01$ therefore provides a good trade off between accuracy and time, as well as ensuring our algorithm successfully returns a shape for each halo.

\section{Convergence of Einstein radii}
\label{App:lensing_tests}

The lensing analysis presented in Section~\ref{sect:lensing} relies on various numerical parameters, which must be sensibly chosen to produce correct results. These parameters include the total area and pixel size of the initial convergence map, the amount of zero padding that we surround this map with when calculating deflection angles (using DFT-based methods) and the density of grid points at which we interpolate $\vect{\alpha}(\vect{x})$, where the interpolation is done from an $\vect{\alpha}(\vect{x})$ grid with the same resolution as the initial convergence map. We experimented with factor of two changes\footnote{Factor of two in lengths, meaning a factor of four in areas.} to all of these quantities and found that our lensing results were unchanged.

While it is reassuring that we have chosen the above-mentioned parameters successfully, this \emph{convergence} with respect to these parameters should not come as a surprise, as our method implicitly smooths the 2D density field through its use of ATSC. This means that the convergence field is smooth on some scale -- set by the density of simulation particles, such that increasing the resolution of the lensing procedure (i.e. decreasing the pixel size of the convergence map, or interpolating $\vect{\alpha}(\vect{x})$ onto a finer grid) will have little effect so long as the resolution of the lensing procedure is already small compared with the scale on which the density field is smooth. The question we would like to answer is whether we would obtain the same results had we run higher-resolution simulations. Although we do not have such simulations, we can degrade the simulations we do have, and see where we recover converged results from these degraded simulations.

\begin{figure}
        \centering
        \includegraphics[width=\columnwidth]{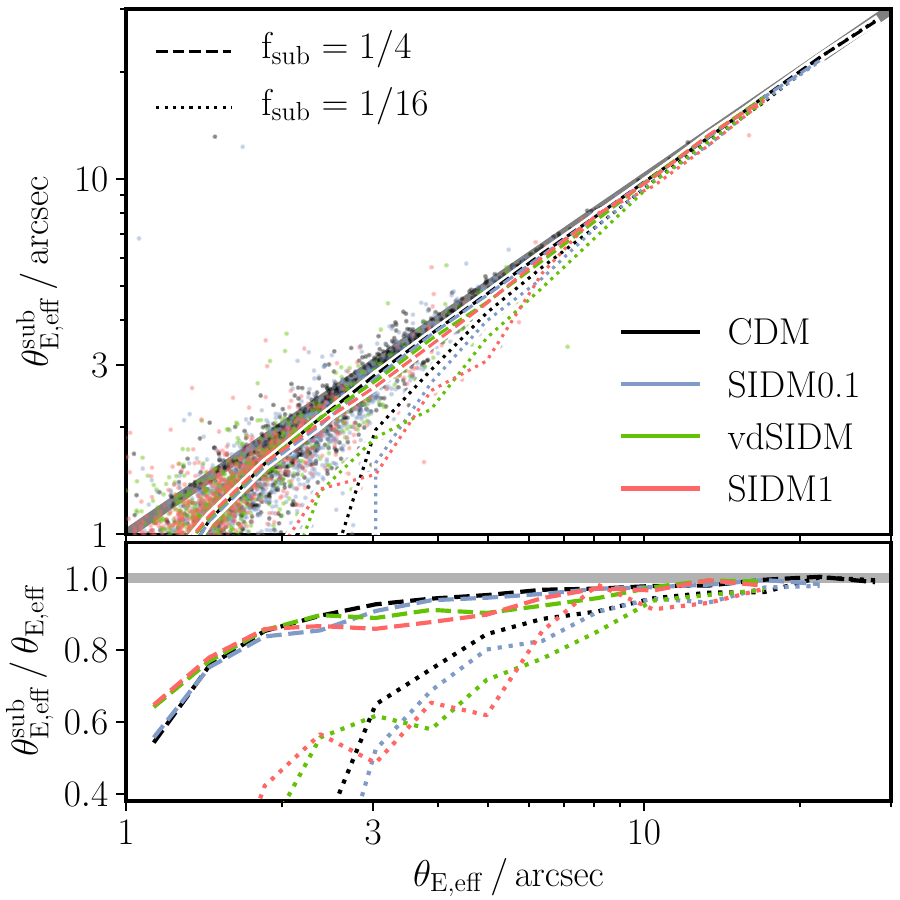}
	\caption{Top: the effective Einstein radius when using only every fourth or sixteenth particle from the simulations, versus the result using all the particles. The points show individual haloes when using every fourth particle, with the dashed lines the median relations (binned by $\theta_\mathrm{E,eff}$ when not subsampling). The dotted lines show the median relations when only a sixteenth of the simulation particles are used. The colours are for the different cross-sections as used throughout the rest of this paper. Bottom: the median lines from the top panel expressed as a relative value, showing the fractional bias in $\theta_\mathrm{E,eff}$ when subsampling.}
	\label{fig:theta_E_convergence}
\end{figure}

We generate lower-resolution simulation data by subsampling the particles from our simulations. We randomly select a fraction, $f_\mathrm{sub}$, of the simulation particles and increase their masses by $1/f_\mathrm{sub}$, throwing away the other particles. We then repeat our lensing analysis using these subsampled simulations, and compare the results with those from the full simulation. Unsurprisingly, the most massive haloes are least sensitive to being subsampled, while results for lower-mass haloes and especially systems with small $\effein$ can vary dramatically after this subsampling procedure. In Fig.~\ref{fig:theta_E_convergence} we show the extent to which $\effein$ is converged with respect to subsampling, showing the results with $f_\mathrm{sub} = 1/4$ and $1/16$. With a lower density of particles, the ATSC scheme smooths the density on a larger physical scale, which generally decreases $\effein$. While the extent to which $\effein$ is converged does not depend solely on $\effein$, this is approximately the case, with the dependence on halo mass or DM model only slight.

We can use the results of Fig.~\ref{fig:theta_E_convergence} to estimate the level to which we can trust the results from our full simulation. The median subsampled $\effein$ (using all DM models) drops to 90\% of that from the full simulation at 6.4 and 2.4 arcsec for $f_\mathrm{sub} = 1/16$ and $1/4$ respectively. The trend in convergence becomes more clear if we look at the $\effein$ at which the median value with $f_\mathrm{sub}=1/16$ drops to 90\% of that with $f_\mathrm{sub}=1/4$. This happens at 5.0 arcsec, implying a factor of $\approx 4$ decrease in the area enclosed by the smallest \emph{converged} critical curves as the density of simulation particles is increased by a factor of four. This suggests that convergence requires a constant number ($\approx 50$) of particles to be enclosed by the critical curves. Given this, we expect that our lensing results from the full simulation are trustworthy down to $\effein \approx 1.2$ arcsec, so we conservatively claim convergence down to 2 arcsec. 

Note that our procedure for generating a convergence map -- using ATSC with the distance to the 8th nearest neighbour used to determine the smoothing-scale of each particle -- was chosen on the basis of having the best convergence characteristics compared with TSC, ATSC with a 32nd nearest neighbour distance and a scheme similar to ATSC but with a Gaussian smoothing kernel. While better schemes for generating 2D mass maps for lensing may exist \citep[for a good overview of methods for converting particles into continuous density fields see][]{density_estimation}, Fig.~\ref{fig:theta_E_convergence} demonstrates that our method is robust, particularly for the most massive haloes in our simulations.

\bsp
\label{lastpage}

\end{document}